\documentstyle[epsfig]{aipproc}
\def\cond {\langle \varphi \rangle}
\newcommand{\be}{\begin{equation}}
\newcommand{\ee}{\end{equation}}
\newcommand{\beq}{\begin{eqnarray}}
\newcommand{\eeq}{\end{eqnarray}}

\begin{document}
\title{Inflation, Topological Defects and Baryogenesis: Selected
     Topics at the Interface between Particles \& Fields and Cosmology\thanks{Invited lectures at the VII Mexican School of Particles and Fields and the I Latin American Symposium on High Energy Physics, Merida, Mexico, 10/29 - 11/6 1996.}}

\author{Robert H. Brandenberger}
\address{Brown University Physics Department\\
Providence, R.I. 02912, USA}

%\lefthead{LEFT head}
%\righthead{RIGHT head}
\maketitle

\begin{abstract}
Modern cosmology provides a connection between the physics of particles and fields and observational cosmology. Making use of this link, a wealth of new observational data can be utilized to explore and constrain theories of fundamental physics. Inflationary Universe models and topological defect theories are the most popular current paradigms for explaining the origin of structure in the Universe. In these lectures, I discuss various aspects of inflation and topological defects in which there has been interesting recent progress or in which there are outstanding problems. Particular emphasis is given to how baryogenesis scenarios can be influenced by inflation and topological defects. 
\end{abstract}

\section*{Introduction and Outline}

Most aspects of high energy physics beyond the standard model can only be tested by going to energies far greater than those which present accelerators can provide. Fortunately, the marriage between particle physics and cosmology has provided a way to ``experimentally" test the new theories of fundamental forces.

The key realization, discovered both in the context of the scenario of inflation$^{\cite{Guth}}$ and of topological defects models$^{\cite{ZelVil}}$ is that physics of the very early Universe may explain the origin of the structure which is observed. It now appears that a rich set of data concerning the nonrandom distribution of matter on a wide range of cosmological scales, and on the anisotropies in the cosmic microwave background (CMB), may potentially be explained by high energy physics. In addition, studying the consequences of particle physics models in the context of cosmology may lead to severe constraints on new microscopic theories. Finally, particle physics and field theory may provide explanations of some deep cosmological puzzles, e.g. why the Universe at the present time appears so homogeneous, so close to being spatially flat, and why it contains the observed small net baryon to entropy ratio.

In these lectures, I focus on three important aspects of modern cosmology. The first concerns some fundamental problems of inflationary cosmology. In particular, some recent progress in the understanding of ``reheating" in inflation will be reviewed.

The second topic concerns topological defect models of structure formation (due to lack of space and time I focus mostly on cosmic strings). Although at the moment defect theories do not explain some of the basic problems of standard cosmology which inflation does, defects do provide conceptually straightforward and in principle quite predictive theories of structure formation. I will review the main points of the cosmic string model, focusing on the predictions with which defect models and inflation-based structure formation theories can be distinguished.

The third main topic is baryogenesis. Recent progress on electroweak baryogenesis will be reviewed, with particular attention to the role which topological defects may play. 

The specific outline is as follows:
\begin{enumerate}
\item{} {\bf Introduction and Outline}
\item{} {\bf Inflationary Universe: Progress and Problems}
\\{2.A} Problems of Standard Cosmology
\\{2.B} Inflationary Universe Scenario
\\{2.C} Problems of Inflation
\\{2.D} Reheating in Inflationary Cosmology
\\{2.E} Summary
\item{} {\bf Topological Defects and Structure Formation}  
\\{3.A} Quantifying Data on Large-Scale Structure
\\{3.B} Topological Defects
\\{3.C} Formation of Defects in Cosmological Phase Transitions
\\{3.D} Evolution of Strings and Scaling
\\{3.E} Cosmic Strings and Structure Formation
\\{3.F} Specific Predictions
\item{} {\bf Topological Defects and Baryogenesis}
\\{4.A} Principles of Baryogenesis
\\{4.B} GUT Baryogenesis and Topological Defects
\\{4.C} Electroweak Baryogenesis and Topological Defects
\\{4.D} Summary
\end{enumerate}

In the Merida lectures I also discussed a further important topic, the classical and quantum theory of cosmological perturbations, which has become the main tool of modern cosmology. A general relativistic and quantum mechanical analysis of the generation and evolution of linearized fluctuations is essential in order to be able to accurately calculate the amplitude of density perturbations and CMB anisotropies. However, due to lack of space I refer the readers interested in this topic to other recent conference contributions$^{\cite{RB97,MFBrev}}$ and to a detailed review article$^{\cite{MFB92}}$.

Unless otherwise specified, units in which $\hbar = c = k_B = 1$ will be used. Distances are expressed in Mpc (1pc $\simeq$ 3.06 light years). Following the usual convention, $h$ indicates the expansion rate of the Universe in units of $100$ km s$^{-1}$ Mpc$^{-1}$, $\Omega = \rho / \rho_c$ is the ratio of the energy density $\rho$ to the critical density $\rho_c$ (the density which yields a spatially flat Universe), $G$ is Newton's constant and $m_{pl}$ is the Planck mass.

\section*{Inflationary Universe: Progress and Problems}

The hypothesis that the Universe underwent a period of exponential expansion at very early times has become the most popular theory of the early Universe. Not only does it solve some of the problems of standard big bang cosmology, but it also provides a causal theory for the origin of inhomogeneities in the Universe which is predictive and in reasonable agreement with current observational results. Nevertheless, there are several problems of principle which merit further study.

\subsection*{Problems of Standard Cosmology}

The standard big bang cosmology rests on three theoretical pillars: the
cosmological principle, Einstein's general theory of relativity and a perfect
fluid description of matter.

The cosmological principle states that on large distance scales the
Universe is homogeneous. This implies that the metric of space-time can be written in Friedmann-Robertson-Walker (FRW) form:
\be
 ds^2 = a(t)^2 \, \left[ {dr^2\over{1-kr^2}} + r^2 (d \vartheta^2 + \sin^2
\vartheta d\varphi^2) \right] \, , %\eqno\eq
\ee
where the constant $k$ determines the topology of the spatial sections. In the following, we shall usually set $k = 0$, i.e. consider a spatially closed Universe. In this case, we can without loss of generality take the scale factor $a(t)$ to be equal to $1$ at the present time $t_0$, i.e. $a(t_0) = 1$. The coordinates $r, \vartheta$ and $\varphi$ are comoving spherical coordinates. World lines with constant comoving coordinates are geodesics corresponding to particles at rest. If the Universe is expanding, i.e. $a(t)$ is increasing, then the physical distance $\Delta x_p(t)$ between two points at rest with fixed comoving distance $\Delta x_c$ grows:
\be
\Delta x_p = a(t) \Delta x_c \, . %\eqno\eq
\ee
 
The dynamics of an expanding Universe  is determined by the Einstein equations,
which relate the expansion  rate to the matter content, specifically to the
energy density $\rho$ and pressure $p$.  For a homogeneous and isotropic
Universe, they reduce to the Friedmann-Robertson-Walker (FRW) equations
\be
\left( {\dot a \over a} \right)^2 - {k\over a^2} = {8 \pi G\over 3 } \rho
\ee
\be
{\ddot a\over a} = - {4 \pi G\over 3} \, (\rho + 3 p) \, .
\ee
These equations can be combined to yield the continuity equation (with Hubble
constant $H = \dot a/a$)
\be \label{cont}
\dot \rho = - 3 H (\rho + p) \, . %\eqno\eq
\ee

The third key assumption of standard cosmology is that matter is described by
an ideal gas with an equation of state
\be
p = w \rho \, . %\eqno\eq
\ee
For cold matter, pressure is negligible and hence $w = 0$.  From (\ref{cont}) it
follows that
\be
\rho_m (t) \sim a^{-3} (t) \, , %\eqno\eq
\ee
where $\rho_m$ is the energy density in cold matter.  For radiation we have $w
= {1/3}$ and hence it follows from (\ref{cont}) that
\be
\rho_r (t) \sim a^{-4} (t) \, , %\eqno\eq
\ee
$\rho_r (t)$ being the energy density in radiation.

The three classic observational pillars of standard cosmology are Hubble's law, the existence and black body nature of the nearly isotropic CMB, and the abundances of light elements (nucleosynthesis). These successes are discussed in detail in many textbooks on cosmology, and will therefore not be reviewed here.

It is, however, important to recall two important aspects concerning the thermal history of the early Universe. Since the energy density in radiation redshifts faster than the matter energy density, it follows by working backwards in time from the present data that although the energy density of the Universe is now mostly in cold matter, it was initially dominated by radiation. The transition occurred at a time denoted by $t_{eq}$, the ``time of equal matter and radiation", which is also the time when perturbations can start to grow by gravitational clustering. The second important time is $t_{rec}$, the ``time of recombination" when photons fell out of equilibrium. The photons of the CMB have travelled without scattering from $t_{rec}$. Their spatial distribution is predicted to be a black body since the cosmological redshift preserves the black body nature of the initial spectrum (simply redshifting the temperature) which was in turn determined by thermal equilibrium. CMB anisotropies probe the density fluctuations at $t_{rec}$. Note that for the usual values of the cosmological parameters, $t_{eq} < t_{rec}$. 

Standard Big Bang cosmology is faced with several important problems.  Only one
of these,  the age problem, is a potential conflict with observations.  The
three problems which are most often discussed in the context of inflation -- the homogeneity, flatness and formation of
structure problems (see e.g. \cite{Guth}) -- are questions which have no answers
within the standard theory but which can be successfully addressed in the context of inflationary cosmology.

The horizon problem is the fact that, within the context of standard cosmology, the comoving
region $\ell_p (t_{rec})$ over which the CMB is observed to be homogeneous  to
better  than one part in $10^4$ is much larger than the comoving forward light
cone $\ell_f (t_{rec})$ at $t_{rec}$, which is the maximal distance over which
microphysical forces could have caused the homogeneity. Hence, standard cosmology cannot explain the observed isotropy of the CMB.
 
In standard cosmology and in an expanding Universe, $\Omega = 1$ is an unstable
fixed point.  As the temperature decreases, $\Omega - 1$ increases.  In fact, in order
to explain the present small value of $\Omega - 1 \sim {\cal O} (1)$, the
initial energy density had to be extremely close to critical density.  For
example, at $T = 10^{15}$ GeV, we require  
$\Omega - 1  \sim 10^{-50}$. What is the origin of these fine tuned initial conditions?  This is the flatness problem of standard cosmology. 

\begin{figure}[b!] % fig 1
\centerline{\epsfig{file=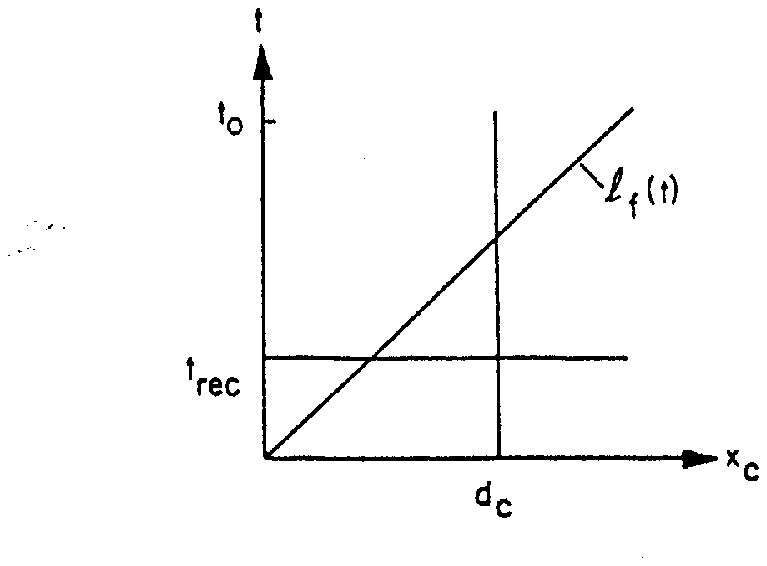,height=2.8in,width=3.5in}}
\vspace{10pt}
\caption{A sketch (conformal separation vs. time) of the formation of
structure problem: the comoving separation $d_c$ between two clusters is larger
than the forward light cone at time $t_{eq}$.}
\label{rhbfig1}
\end{figure}

The third of the classic problems of standard cosmological model is the
``formation of structure problem."  Observations indicate that galaxies and
even clusters of galaxies have nonrandom correlations on scales larger than 50
Mpc (see e.g. \cite{CFA,LCRS}).  This scale is comparable to the comoving horizon at
$t_{eq}$.  Thus, if the initial density perturbations were produced much before
$t_{eq}$, the correlations cannot be explained by a causal mechanism.  Gravity
alone is, in general, too weak to build up correlations on the scale of
clusters after $t_{eq}$ (see, however, the explosion scenario of \cite{explosion}).
Hence, the two questions of what generates the primordial density perturbations
and what causes the observed correlations, do not have an answer in the context
of standard cosmology.  This problem is illustrated in Fig. \ref{rhbfig1}.

There are other serious problems of standard cosmology, e.g. the age and the cosmological constant problems. However, to date modern cosmology does not shed any light on these problems, and I will therefore not address them here.

\subsection*{Inflationary Universe Scenario}

The idea of inflation$^{\cite{Guth}}$ is very simple (for some early reviews of inflation see e.g. \cite{Linde,GuthBlau,Olive,RB85}).  We assume there is a time
interval beginning at $t_i$ and ending at $t_R$ (the ``reheating time") during
which the Universe is exponentially expanding, i.e.,
\be
a (t) \sim e^{Ht}, \>\>\>\>\> t \epsilon \, [ t_i , \, t_R] 
\ee
with constant Hubble expansion parameter $H$.  Such a period is called  ``de
Sitter" or ``inflationary."  The success of Big Bang nucleosynthesis sets an
upper limit to the time of reheating, namely the time of nucleosynthesis.

\begin{figure} [b!]
\centerline{\epsfig{file=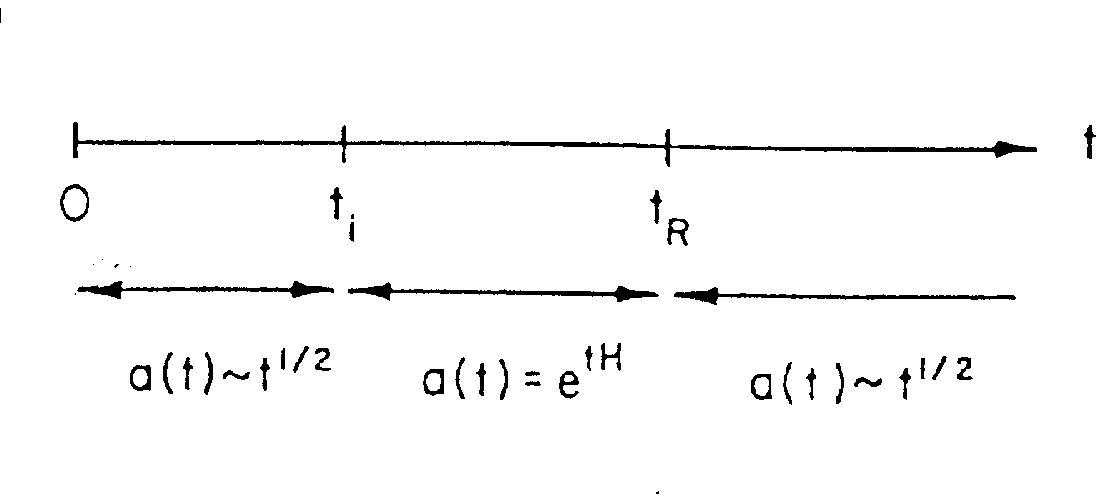,height=2in,width=3.5in}}
\vspace{10pt}
\caption{The phases of an inflationary Universe. The times
$t_i$ and $t_R$ denote the beginning and end of inflation, respectively.
In some models of inflation, there is no initial radiation dominated FRW
period. Rather, the classical space-time emerges directly in an inflationary
state from some initial quantum gravity state.}
\label{rhbfig2}
\end{figure}  

The phases of an inflationary Universe are sketched in Fig. \ref{rhbfig2}.  Before the
onset of inflation there are no constraints on the state of the Universe.  In
some models a classical space-time emerges immediately in an inflationary
state, in others there is an initial radiation dominated FRW period.  Our
sketch applies to the second case.  After $t_R$, the Universe is very hot and
dense, and the subsequent evolution is as in standard cosmology.  During the
inflationary phase, the number density of any particles initially in thermal
equilibrium at $t = t_i$ decays exponentially.  Hence, the matter temperature
$T_m (t)$ also decays exponentially.  At $t = t_R$, all of the energy which is
responsible for inflation (see later) is released as thermal energy.  This is a
nonadiabatic process during which the entropy increases by a large factor.   

Inflation can easily solve the homogeneity
problem. Let $\Delta t = t_R - t_i$  denote the period of inflation.  During
inflation, the forward light cone increases exponentially compared to a model
without inflation, whereas the past light cone is not affected for $t \geq
t_R$.  Hence, provided $\Delta t$ is sufficiently large, $\ell_f (t_R)$ will be
greater than $\ell_p (t_R)$.  

Inflation also can solve the flatness problem$^{\cite{Kazanas,Guth}}$  The key point is
that the entropy density $s$ is no longer constant.  As will be explained
later, the temperatures at $t_i$ and $t_R$ are essentially equal.  Hence, the
entropy increases during inflation by a factor $\exp (3 H \Delta t)$.  Thus,
$\epsilon$ decreases by a factor of $\exp (-2 H \Delta t)$.  Hence, $(\rho - \rho_c) / \rho$ can be of order 1 both at $t_i$ and at the
present time.  In fact, if inflation occurs at all, then rather generically, the theory predicts
that at the present time $\Omega = 1$ to a high accuracy (now $\Omega < 1$
requires  special initial conditions or rather special models$^{\cite{open}}$).

\begin{figure}[b!]
\centerline{\epsfig{file=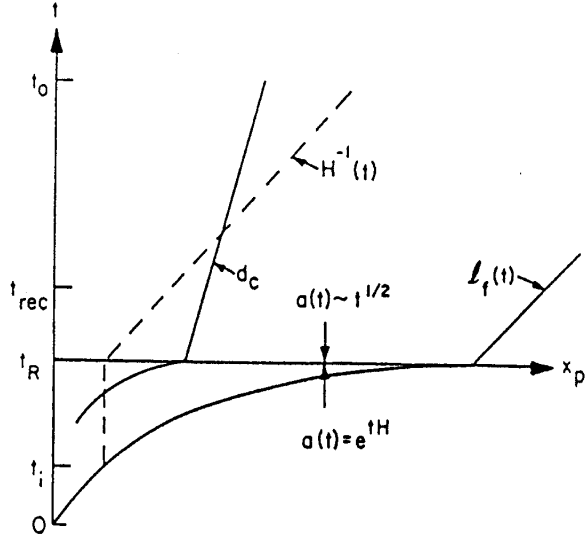,height=3.5in,width=3.5in}}
\vspace{10pt}
\caption{A sketch (physical coordinates vs. time) of the
solution of the formation of structure problem. Provided that the period of
inflation is sufficiently long, the separation $d_c$ between two galaxy
clusters is at all times smaller than the forward light cone. The dashed line
indicates the Hubble radius. Note that $d_c$ starts out smaller than the Hubble
radius, crosses it during the de Sitter period, and then reenters it at late
times.}
\label{rhbfig3}
\end{figure}

Most importantly, inflation provides a mechanism which in a causal way
generates the primordial perturbations required for galaxies, clusters and even
larger objects.  In inflationary Universe models, the Hubble radius
(``apparent" horizon), $3t$, and the ``actual" horizon (the forward light cone)
do not coincide at late times.  Provided that the duration of inflation is sufficiently long, then (as sketched in 
Fig. \ref{rhbfig3}) all scales within our apparent horizon were inside the actual
horizon since $t_i$.  Thus, it is in principle possible to have a casual
generation mechanism for perturbations$^{\cite{Press,Mukh80,Lukash,Sato}}$.

The generation of perturbations is due to a causal microphysical
process.  Such a process can only act coherently on length scales smaller than
the Hubble radius $\ell_H (t)$ where 
$\ell_H (t) = H^{-1} (t)$. A heuristic way to understand the meaning of $\ell_H (t)$ is to realize that it
is the distance which light (and hence the maximal distance any causal effects)
can propagate in one expansion time.

The density perturbations produced during
inflation are due to quantum fluctuations in the matter and gravitational
fields$^{\cite{Mukh80,Lukash}}$.  The amplitude of these inhomogeneities corresponds to a temperature $T_H$ whose value is
$T_H \sim H $, the Hawking temperature of the de Sitter phase. This implies that at all times
$t$ during inflation, perturbations with a fixed physical wavelength $\sim
H^{-1}$ will be produced. Subsequently, the length of the waves is stretched
with the expansion of space, and soon becomes larger than the Hubble radius.
The phases of the inhomogeneities are random.  Thus, the inflationary Universe
scenario predicts perturbations on all scales ranging from the comoving Hubble
radius at the beginning of inflation to the corresponding quantity at the time
of reheating.  In particular, provided that inflation lasts sufficiently long, perturbations on scales of galaxies and beyond will be generated. Note, however, that it is very dangerous to interpret de Sitter Hawking radiation as thermal radiation. In fact, the equation of state of this ``radiation" is not thermal$^{\cite{RB83}}$.

Obviously, the key question is how to obtain inflation. From the FRW equations, it follows that in order to get exponential increase of the scale factor, the equation of state of matter must be
\be \label{infleos}
p = - \rho  
\ee
This is where the connection with particle physics comes in. The energy density and pressure of a scalar quantum field $\varphi$ are given by
\beq
\rho (\varphi) & = & {1\over 2} \, \dot \varphi^2 + {1\over 2} \,
(\nabla \varphi)^2 + V (\varphi) \label{eos1} \\
p (\varphi) & = & {1\over 2} \dot \varphi^2 - {1\over 6} (\nabla \varphi)^2 - V
(\varphi) \, . \label{eos2}
\eeq
Thus, provided that at some initial time $t_i$
\be \label{incond}
|\dot \varphi (\underline{x}, \, t_i)| , |\nabla \varphi (\underline{x}_i \, t_i)| \ll V (\varphi (\underline{x}_i, \, t_i) ) \, , 
\ee
the equation of state of matter will be (\ref{infleos}).
 
The next question is how to realize the required initial conditions (\ref{incond}) and to maintain the  
dominance of potential over kinetic and gradient energy for sufficiently long. Various ways of realizing these conditions were put forward, and they gave rise to different models of inflation. I will focus on ``old inflation," ``new inflation"" and ``chaotic
inflation."  There are many other attempts at producing an inflationary
scenario, but there is as of now no convincing realization.

\medskip
\centerline{\bf Old Inflation}
\medskip

\begin{figure}
\centerline{\epsfig{file=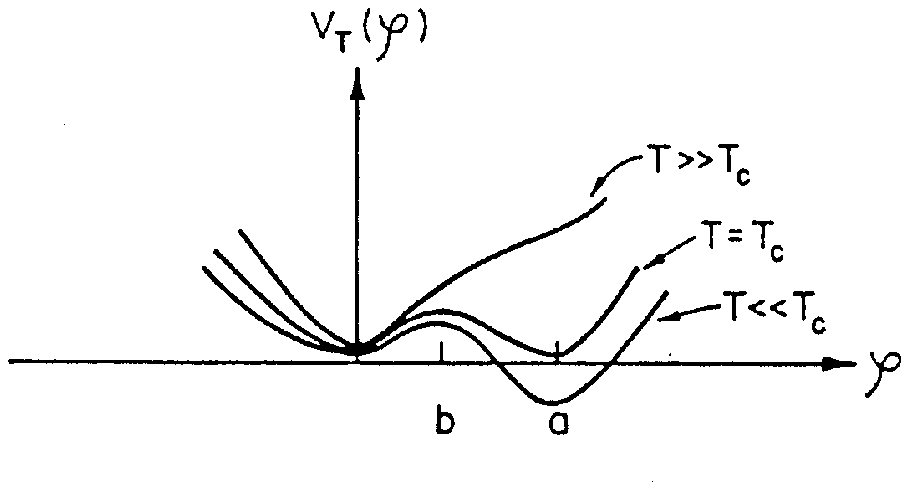,height=2.5in,width=3.5in}}
\caption{The finite temperature effective potential
in a theory with a first order phase transition.}
\label{rhbfig4}
\end{figure}

The old inflationary Universe model$^{\cite{Guth,GuthTye}}$ is based on a scalar field
theory which undergoes a first order phase transition.  As a toy
model, consider a scalar field theory with the potential $V (\varphi)$
of Fig. \ref{rhbfig4}.  This potential has a metastable ``false" vacuum at $\varphi = 0$, whereas the lowest energy state (the ``true" vacuum) is $\varphi = a$. Finite temperature effects$^{\cite{finiteT}}$ lead to extra terms in the finite temperature effective potential which are proportional to $\varphi^2 T^2$ (the resulting finite temperature effective potential is also depicted in Fig. \ref{rhbfig4}). Thus, at high temperatures, the energetically preferred state is the false vacuum state. Note that this is only true if $\varphi$ is in thermal equilibrium with the other fields in the system.

For fairly general initial conditions, $\varphi (x)$ is trapped in the
metastable state $\varphi = 0$ as the Universe cools below the
critical temperature $T_c$.  As the Universe expands further, all
contributions to the energy-momentum tensor $T_{\mu \nu}$ except for
the contribution
\be
T_{\mu \nu} \sim V(\varphi) g_{\mu \nu}  
\ee
redshift.  Hence, provided that the potential $V(\varphi)$ is shifted upwards such that $V(a) = 0$, then the equation of state in the false vacuum approaches $p = - \rho$, and
inflation sets in. After a period $\Gamma^{-1}$, where $\Gamma$ is the tunnelling rate, bubbles of $\varphi = a$ begin to nucleate$^{\cite{decay}}$ in a sea of false
vacuum $\varphi = 0$. Inflation lasts until the false vacuum decays.
During inflation, the Hubble constant is given by
\be
H^2 = {8 \pi G\over 3} \, V (0) \, .  
\ee
Note that the condition $V(a) = 0$, which looks rather unnatural, is required to
avoid a large cosmological constant today (none of the present inflationary Universe
models manages to circumvent or solve the cosmological constant problem).
 
It was immediately realized that old inflation has a serious ``graceful exit"
problem$^{\cite{Guth,GuthWein}}$.  The bubbles nucleate after inflation with radius $r \ll
2t_R$ and would today be much smaller than our apparent horizon.  Thus, unless
bubbles percolate, the model predicts extremely large inhomogeneities inside
the Hubble radius, in contradiction with the observed isotropy of the
microwave background radiation.
\par
For bubbles to percolate, a sufficiently large number must be produced so that
they collide and homogenize over a scale larger than the present Hubble
radius.  However, with exponential expansion, the volume between bubbles
expands
exponentially whereas the volume inside bubbles expands only with a low power.
This prevents percolation.

\medskip
\centerline{\bf New Inflation}
\medskip

Because of the graceful exit problem, old inflation never was considered to be
a viable cosmological model.  However, soon after the seminal paper by
Guth, Linde$^{\cite{Linde82}}$ and independently Albrecht and Steinhardt$^{\cite{AS82}}$ put
forwards a modified scenario, the New Inflationary Universe.

\begin{figure}
\centerline{\epsfig{file=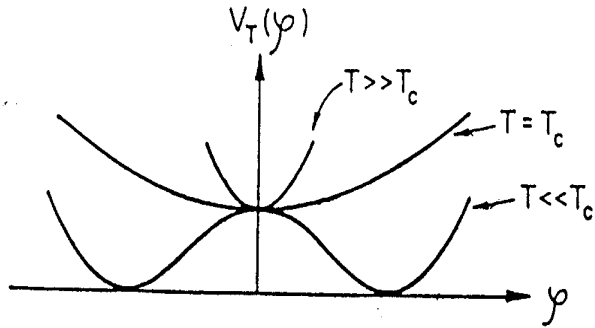,height=2.5in,width=3.5in}}
\caption{ 
The finite temperature effective potential
in a theory with a second order phase transition.}
\label{rhbfig5}
\end{figure}

The starting point is a scalar field theory with a double well potential which
undergoes a second order phase transition (Fig. \ref{rhbfig5}).  $V(\varphi)$ is
symmetric and $\varphi = 0$ is a local maximum of the zero temperature
potential.  Once again, it was argued that finite temperature effects confine
$\varphi(x)$ to values near $\varphi = 0$ at temperatures $T \geq
T_c$.  For $T < T_c$, thermal fluctuations trigger the instability of $\varphi
(x) = 0$ and $\varphi (x)$ evolves towards either of the global minima at $\varphi = \pm \sigma$ by the classical equation of motion
\be \label{eom}
\ddot \varphi + 3 H \dot \varphi - a^{-2} \bigtriangledown^2 \varphi = -
V^\prime (\varphi)\, .  
\ee

Within a fluctuation region, 
$\varphi(x)$ will be homogeneous. In such a region, we can  neglect the spatial gradient terms in Eq. (\ref{eom}).  Then, from (\ref{eos1}) and (\ref{eos2}) we can read off the
induced equation of state.  The condition for inflation is
\be
\dot \varphi^2 \ll V (\varphi)\, , 
\ee
i.e.~ slow rolling.
Often, the  ``slow rolling" approximation is made to find solutions of
(\ref{eom}).
This consists of dropping the $\ddot \varphi$ term.   

There is no graceful exit problem in the new inflationary Universe.  Since the
fluctuation domains are established before the onset of inflation,
any boundary walls will be inflated outside the present Hubble radius.
\par
Let us, for the moment, return to the general features of the new inflationary
Universe scenario.  At the time $t_c$ of the phase transition, $\varphi (t)$
will start to move from near $\varphi = 0$ towards either $\pm \sigma$ as
described by the classical equation of motion, i.e.~ (\ref{eom}).  At or soon after
$t_c$, the energy-momentum tensor of the Universe will start to be dominated
by $V(\varphi)$, and inflation will commence.  $t_i$ shall denote the time of
the onset of inflation.  Eventually, $\phi (t)$ will reach large values for which
nonlinear effects become important.  The time at which this occurs is $t_B$.
For $t > t_B \, , \, \varphi (t)$ rapidly accelerates, reaches $\pm \sigma$,
overshoots and starts oscillating about the global minimum of $V (\varphi)$.
The amplitude of this oscillation is damped by the expansion of the Universe
and (predominantly) by the coupling of $\varphi$ to other fields.  At time
$t_R$,
the energy in $\varphi$ drops below the energy of the thermal bath of
particles produced during the period of oscillation.
\par
\begin{figure}
\centerline{\epsfig{file=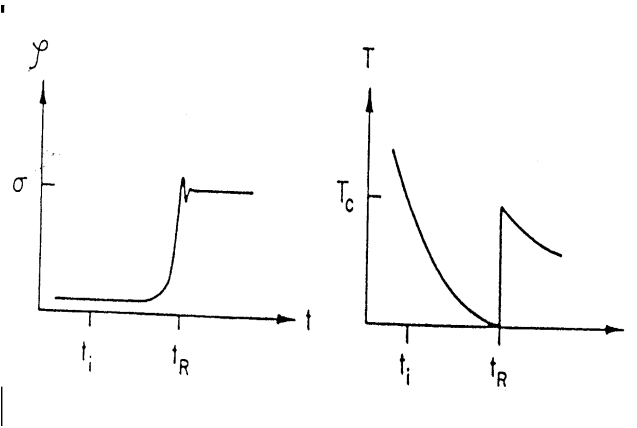,height=2.2in,width=3.5in}}
\caption{Evolution of $\varphi (t)$ and $T (t)$ in the new
inflationary Universe.}
\label{rhbfig6}
\end{figure}

The evolution of $\varphi (t)$ is sketched in Fig. \ref{rhbfig6}.  The time period
between $t_B$ and $t_R$ is called the reheating period and is usually short
compared to the Hubble expansion time. For $t > t_R$, the Universe is again radiation dominated.

In order to obtain inflation, the potential $V(\varphi)$ must be very flat near the false vacuum at $\varphi = 0$. This can only be the case if all of the coupling constants appearing in the potential are small. However, this implies that the  $\varphi$ cannot be in thermal equilibrium at early times, which would be required to localize $\varphi$ in the false vacuum. In the absence of thermal equilibrium, the initial conditions for $\varphi$ are only constrained by requiring that the total energy density in $\varphi$ not exceed the total energy density of the Universe. Most of the phase space of these initial conditions lies at values of $| \varphi | >> \sigma$. This leads to the ``chaotic" inflation scenario$^{\cite{Linde83}}$.

\medskip  
\centerline{\bf Chaotic Inflation}
\medskip

Consider a region in space where at the initial time $\varphi (x)$
is very large, homogeneous and static.  In this case, the energy-momentum tensor will be
immediately dominated by the large potential energy term and induce an
equation of state $p \simeq - \rho$ which leads to inflation.  Due to the
large Hubble damping term in the scalar field equation of motion, $\varphi
(x)$ will only roll very slowly towards $\varphi = 0$.  The
kinetic energy contribution to $T_{\mu \nu}$ will remain small, the spatial
gradient contribution will be exponentially suppressed due to the expansion of
the Universe, and thus inflation persists. Note that in contrast to old and new inflation,
no initial thermal bath is required.  Note also that the precise form of
$V(\varphi)$ is irrelevant to the mechanism.  In particular, $V(\varphi)$ need
not be a double well potential.  This is a significant advantage, since for
scalar fields other than Higgs fields used for spontaneous symmetry breaking,
there is no particle physics motivation for assuming a double well potential,
and since the inflaton (the field which gives rise to inflation) cannot be a
conventional Higgs field due to the severe fine tuning constraints.
\par
The field and temperature evolution in a chaotic inflation model is similar to what is depicted in Fig. \ref{rhbfig6}, except that $\varphi$ is rolling towards the true vacuum at $\varphi = \sigma$ from the direction of large field values.

Chaotic inflation is a much more radical departure from standard cosmology than old and new inflation. In the latter, the inflationary phase can be viewed as a short phase of exponential expansion bounded at both ends by phases of radiation domination. In chaotic inflation, a piece of the Universe emerges with an inflationary equation of state immediately after the quantum gravity (or string) epoch.

The chaotic inflationary Universe scenario has been developed in great detail (see e.g. \cite{Linde94} for a recent review). One important addition is the inclusion of stochastic noise$^{\cite{Starob87}}$ in the equation of motion for $\varphi$ in order to take into account the effects of quantum fluctuations. It can in fact be shown that for sufficiently large values of $|\varphi|$, the stochastic force terms are more important than the classical relaxation force $V^\prime(\varphi)$. There is equal probability for the quantum fluctuations to lead to an increase or decrease of $|\varphi|$. Hence, in a substantial fraction of comoving volume, the field $\varphi$ will climb up the potential. This leads to the conclusion that chaotic inflation is eternal. At all times, a large fraction of the physical space will be inflating. Another consequence of including stochastic terms is that on large scales (much larger than the present Hubble radius), the Universe will look extremely inhomogeneous.

\subsection*{Problems of Inflationary Cosmology}

In spite of its great success at resolving some of the problems of standard cosmology and of providing a causal, predictive theory of structure formation, there are several important unresolved conceptual problems in inflationary cosmology. I will focus on three of these problems, the cosmological constant mystery, the fluctuation problem, and the dynamics of reheating.

\medskip
\centerline{\bf Cosmological Constant Problem}
\medskip

Since the cosmological constant acts as an effective energy density, its value is bounded from above by the present energy density of the Universe. In Planck units, the constraint on the effective cosmological constant $\Lambda_{eff}$ is
(see e.g. \cite{cosmorev})
\be
{{\Lambda_{eff}} \over {m_{pl}^4}} \, \le \, 10^{- 122} \, .
\ee
This constraint applies both to the bare cosmological constant and to any matter contribution which acts as an effective cosmological constant.

The true vacuum value of the potential $V(\varphi)$ acts as an effective cosmological constant. Its value is not constrained by any particle physics requirements (in the absence of special symmetries). The cosmological constant problem is thus even more accute in inflationary cosmology than it usually is. The same unknown mechanism which must act to shift the potential (see Fig. \ref{rhbfig4}) such that inflation occurs in the false vacuum must also adjust the potential to vanish in the true vacuum. 

Supersymmetric theories may provide a resolution of this problem, since unbroken supersymmetry forces $V(\varphi) = 0$ in the supersymmetric vacuum. However, supersymmetry breaking will induce a nonvanishing $V(\varphi)$ in the true vacuum after supersymmetry breaking.

We may therefore be forced to look for realizations of inflation which do not make use of scalar fields. There are several possibilities. It is possible to obtain inflation in higher derivative gravity theories. In fact, the first
model with exponential expansion of the Universe was obtained$^{\cite{Starob}}$ in an $R^2$ gravity theory. The extra degrees of freedom associated with the higher derivative terms act as scalar fields with a potential which automatically vanishes in the true vacuum. For some recent work on higher derivative gravity inflation see also \cite{MB92}. 

Another way to obtain inflation is by making use of condensates (see \cite{Ball} and \cite{Parker} for different approaches to this problem). An additional motivation for following this route to inflation is that the symmetry breaking mechanisms observed in nature (in condensed matter systems) are induced by the formation of condensates such as Cooper pairs. Again, in a model of condensates there is no freedom to add a constant to the effective potential.

The main problem when studying the possibility of obtaining inflation using condensates is that the quantum effects which determine the theory are highly nonperturbative. In particular, the effective potential written in terms of a condensate $\cond$ does not correspond to a renormalizable theory and will in general$^{\cite{ARZ}}$ contain terms of arbitrary power in $\cond$. However (see \cite{BZ96}), one may make progress by assuming certain general properties of the effective potential.

Let us$^{\cite{BZ96}}$ consider a theory in which at some time $t_i$ a condensate $\cond$ forms, i.e. $\cond = 0$ for $t < t_i$ and $\cond \neq 0$ for $t > t_i$. The expectation value of the Hamiltonian $H$ written in terms of the condensate $\cond$ contains terms of arbitrary powers of $\cond$:
\be
\langle H \rangle \, = \, \sum_{n} (-1)^n {n!} a_n \cond^n \, .
\ee
We summarize our ignorance of the nonperturbative physics in the assumption that the resulting series is asymptotic, and in particular Borel summable, with coefficients $a_n \propto 1$. In this case, we can resum the series to obtain$^{\cite{BZ96}}$
\be \label{effpot}
\langle H \rangle \, = \, \int_0^{\infty} {{f(t) dt} \over {t (t m_{pl} + \cond)}} e^{- 1/t} \, ,
\ee
where the function $f(t)$ is related to the coefficients $a_n$ via
\be
a_n \, = \, {1 \over {n!}} \int_0^{\infty} dt f(t) t^{-n - 2} e^{- 1/t} \, .
\ee

The expectation value of the Hamiltonian $\langle H \rangle$ can be interpreted as the effective potential $V_{eff}$ of this theory. The question is under which conditions this potential gives rise to inflation. If we regard $\cond$ as a classical field (i.e. neglect the ultraviolet and infrared divergences of the theory), then the dynamics of the model can be read off directly from (\ref{effpot}), with initial conditions for $\cond$ at the time $t_i$ close to $\cond = 0$. It is easy to check that rather generically, the conditions required to have slow rolling of $\varphi$, namely
\be
V^\prime m_{pl} \, << \, \sqrt{48 \pi} V 
\ee
\be
V^{\prime \prime} m_{pl}^2 \, << \, 24 \pi V \, ,
\ee
are satisfied. However, since the potential decays only slowly at large values of $\cond$ and since there is no true vacuum state at finite values of $\cond$, the slow rolling conditions are satisfied for all times. In this case, inflation would never end - an obvious cosmological disaster.

However, $\cond$ is not a classical scalar field but the expectation value of a condensate operator. Thus, we have to worry about diverging contributions to this expectation value. In particular, in a theory with symmetry breaking there will often be massless excitations which will give rise to infrared divergences. It is necessary to introduce an infrared cutoff energy $\varepsilon$ whose value is determined in the context of cosmology by the Hubble expansion rate. Note in particular that this cutoff is time-dependent. Effectively, we thus have a theory of two scalar fields $\cond$ and $\varepsilon$. 
In this case, the first of the slow rolling conditions becomes (if $\varepsilon$ is expressed in Planck units)
\be
\dot{\varepsilon}^2 m_{pl}^2 + \dot{\varphi}^2 \, << \, 2 V \, .
\ee

The infrared cutoff changes the form of the effective potential. We assume that this change can be modelled by replacing $\cond$ by $\cond / \varepsilon$. If we
(following \cite{Woodard}) take the infrared cutoff to be
\be \label{ansatz}
\varepsilon(t) = {{H(0)} \over {m_{pl}}} [1 - a (Ht)^p] \, ,
\ee
where $0 < a << 1$ and $p$ is an integer and the time at the beginning of the rolling has been set to $t = 0$, then it can be shown$^{\cite{BZ96}}$ that an period of inflation with a graceful exit is realized. After the
condensate $\cond$ starts rolling at $\cond \sim 0$, inflation will commence. As inflation proceeds, $\varepsilon(t)$ will slowly grow and will eventually dominate the energy functional, signaling an end of the inflationary period. From (\ref{ansatz}) it follows that inflation lasts until $a^{1/p} H t = 1$. 

This analysis demonstrates that it is in principle possible to obtain inflation from condensates. However, the model must be studied in much more detail before we can determine whether it gives a realization of inflation which is free of problems.

\medskip
\centerline{\bf Fluctuation Problem}
\medskip

A generic problem for all realizations of inflation studied up to now concerns the amplitude of the density perturbations which are induced by quantum fluctuations during the period of exponential expansion. From the amplitude of CMB anisotropies measured by COBE, and from the present amplitude of density inhomogeneities on scales of clusters of galaxies, it follows that the amplitude of the mass fluctuations ${\delta M} / M$ on a length scale given by the comoving wavenumber $k$ at the time $t_H(k)$ when that scale crosses the Hubble radius in the FRW period is 
\be \label{obs}
{{\delta M} \over M} (k, t_H(k)) \, \propto \, 10^{-5} \, .
\ee

The perturbations originate during inflation as quantum excitations (see e.g. \cite{MFB92} for a comprehensive review). Their amplitude at the time $t_i(k)$ when the scale $k$ leaves the Hubble radius during inflation is given by
\be \label{inpert}
{{\delta M} \over M} (k, t_i(k)) \, \simeq \, {{V^{\prime} \delta \varphi} \over \rho}|_{t_i(k)} \, ,
\ee
where $\delta \varphi$ is the amplitude of the quantum fluctuation of $\delta \varphi(k)$ (note that this is a momentum space quantity). While the scale $k$ is larger than the Hubble radius, the fluctuation amplitude grows by general relativistic gravitational effects. The amplitudes at $t_i(k)$ and $t_H(k)$ are related by
\be \label{amplpert}
{{\delta M} \over M} (k, t_H(k)) \, \simeq \, {1 \over {1 + p / \rho}}|_{t_i(k)}
{{\delta M} \over M} (k, t_i(k)) 
\ee
(see e.g. \cite{BST,BK84,MFB92}). Combining (\ref{inpert}) and (\ref{amplpert}) and working out the result for the potential
\be
V(\varphi) \, = \, \lambda \varphi^4
\ee
we obtain the result$^{\cite{Mukh81,flucts,BST}}$
\be
{{\delta M} \over M} (k, t_H(k)) \, \simeq 10^2 \lambda^{1/2} \, .
\ee
Thus, in order to agree with the observed value (\ref{obs}), the coupling constant $\lambda$ must be extremely small:
\be \label{fluctconstr}
\lambda \, \leq \, 10^{-12} \, .
\ee

It has been shown in \cite{Freese} that the above conclusion is generic, at least for models in which inflation is driven by a scalar field. In order that inflation does not produce a too large amplitude of the spectrum of perturbations, a dimensionless number appearing in the potential must be set to a very small value. Models in which inflation is NOT driven by a scalar field but realized in some unified theory of all fundamental forces might avoid the fluctuation problem, in particular if there is some principle such as asymptotic freedom during the period of inflation$^{\cite{BMS93}}$ which suppresses scalar perturbations.

\medskip
\centerline{\bf Reheating Problem}
\medskip

A question which has recently received a lot of attention and will be discussed in greater detail shortly is the issue of reheating in inflationary cosmology. The question concerns the energy transfer between the inflaton and matter fields which is supposed to take place at the end of inflation (see Fig. \ref{rhbfig6}). 

According to either new inflation or chaotic inflation, the dynamics of the inflaton leads first to a transfer of energy from potential energy of the inflaton to kinetic energy. After the period of slow rolling, the inflaton $\varphi$ begins to oscillate about the true minimum of $V(\varphi)$. Quantum mechanically, the state of homogeneous oscillation corresponds to a coherent state. Any coupling of $\varphi$ to other fields (and even self coupling terms of $\varphi$) will lead to a decay of this state. This corresponds to the particle production. The produced particles will be relativistic, and thus at the conclusion of the reheating period a radiation dominated Universe will emerge.

The key questions are by what mechanism and how fast the decay of the coherent state takes place. It is important to determine the temperature of the produced particles at the end of the reheating period. The answers are relevant to many important questions regarding the post-inflationary evolution. For example, it is important to know whether the temperature after reheating is high enough to allow GUT baryogenesis and the production of GUT-scale topological defects. In supersymmetric models, the answer determines the predicted abundance of gravitinos and other moduli fields.

Recently, there has been a complete change in our understanding of reheating. This topic will be discussed in detail below.

\subsection*{Reheating in Inflationary Cosmology} 

Reheating is an important stage in inflationary cosmology. It determines the state of the Universe after inflation and has consequences for baryogenesis, defect formation, and, as will be shown below, maybe even for the composition
of the dark matter of the Universe.

After slow rolling, the inflaton field begins to oscillate uniformly in space about the true vacuum state. Quantum mechanically, this corresponds to a coherent state of $k = 0$ inflaton particles. Due to interactions of the inflaton with itself and with other fields, the coherent state will decay into quanta of elementary particles. This corresponds to post-inflationary particle production.

Reheating is usually studied using simple scalar field toy models. The one we will adopt here consists of two real scalar fields, the inflaton $\varphi$
with Lagrangian
\be
{\cal L}_o \, = \, {1 \over 2} \partial_\mu \varphi \partial^\mu \varphi - {1 \over 4} \lambda (\varphi^2 - \sigma^2)^2 
\ee
interacting with a massless scalar field $\chi$ representing ordinary matter. The interaction Lagrangian is taken to be
\be
{\cal L}_I \, = \, {1 \over 2} g^2 \varphi^2 \chi^2 \, .
\ee
Self interactions of $\chi$ are neglected. 

By a change of variables
\be
\varphi \, = \, {\tilde \varphi} + \sigma \, ,
\ee
the interaction Lagrangian can be written as
\be \label{intlag}
{\cal L}_I \, = \, g^2 \sigma {\tilde \varphi} \chi^2 + {1 \over 2} g^2 {\tilde \varphi}^2 \chi^2 \, .
\ee
During the phase of coherent oscillations, the field ${\tilde \varphi}$ oscillates with a frequency
\be
\omega \, = \, m_{\varphi} \, = \, \lambda^{1/2} \sigma 
\ee
(neglecting the expansion of the Universe which can be taken into account as in \cite{KLS94,STB95}).

\medskip
\centerline{\bf Elementary Theory of Reheating}
\medskip

According to the elementary theory of reheating (see e.g. \cite{DolLin} and \cite{AFW}), the decay of the inflaton is calculated using first order perturbations theory. According to the Feynman rules, the decay rate $\Gamma_B$ of $\varphi$ (calculated assuming that the cubic coupling term dominates) is
given by
\be
\Gamma_B \, = \, {{g^2 \sigma^2} \over {8 \pi m_{\phi}}} \, .
\ee
The decay leads to a decrease in the amplitude of $\varphi$ (from now on we will drop the tilde sign) which can be approximated by adding an extra damping term to the equation of motion for $\varphi$:
\be
{\ddot \varphi} + 3 H {\dot \varphi} + \Gamma_B {\dot \varphi} \, = \,
- V^\prime(\varphi) \, .
\ee
From the above equation it follows that as long as $H > \Gamma_B$, particle production is negligible. During the phase of coherent oscillation of $\varphi$, the energy density and hence $H$ are decreasing. Thus, eventually $H = \Gamma_B$, and at that point reheating occurs (the remaining energy density in $\varphi$ is very quickly transferred to $\chi$ particles).

The temperature $T_R$ at the completion of reheating can be estimated by computing the temperature of radiation corresponding to the value of $H$ at which $H = \Gamma_B$. From the FRW equations it follows that
\be
T_R \, \sim \, (\Gamma_B m_{pl})^{1/2} \, .
\ee
If we now use the ``naturalness" constraint{\footnote{At one loop order, the cubic interaction term will contribute to $\lambda$ by an amout $\Delta \lambda \sim g^2$. A renormalized value of $\lambda$ smaller than $g^2$ needs to be finely tuned at each order in perturbation theory, which is ``unnatural".}} 
\be
g^2 \, \sim \, \lambda
\ee
in conjunction with the constraint on the value of $\lambda$ from (\ref{fluctconstr}), it follows that for $\sigma < m_{pl}$,
\be
T_R \, < \, 10^{10} {\rm GeV} \, .
\ee
This would imply no GUT baryogenesis, no GUT-scale defect production, and no gravitino problems in supersymmetric models with $m_{3/2} > T_R$, where $m_{3/2}$ is the gravitino mass. As we shall see, these conclusions change radically if we adopt an improved analysis of reheating.

\medskip
\centerline{\bf Modern Theory of Reheating}
\medskip

However, as was first realized in \cite{TB90}, the above analysis misses an essential point. To see this, we focus on the equation of motion for the matter field $\chi$ coupled to the inflaton $\varphi$ via the interaction Lagrangian ${\cal L}_I$ of (\ref{intlag}). Taking into account for the moment only the cubic interaction term, the equation of motion becomes
\be
{\ddot \chi} + 3H{\dot \chi} - \bigl(({{\nabla} \over a})^2 - m_{\chi}^2 - 2g^2\sigma\varphi \bigr)\chi \, = \, 0 \, .
\ee
Since the equation is linear in $\chi$, the equations for the Fourier modes $\chi_k$ decouple:
\be \label{reseq}
{\ddot \chi_k} + 3H{\dot \chi_k} + (k_p^2 + m_{\chi}^2 + 2g^2\sigma\varphi)\chi_k \, = \, 0 ,
\ee
where $k_p$ is the time-dependent physical wavenumber. 

Let us for the moment neglect the expansion of the Universe. In this case, the friction term in (\ref{reseq}) drops out and $k_p$ is time-independent, and Equation (\ref{reseq}) becomes a harmonic oscillator equation with a time-dependent mass determined by the dynamics of $\varphi$. In the reheating phase, $\varphi$ is undergoing oscillations. Thus, the mass in (\ref{reseq}) is varying periodically. In the mathematics literature, this equation is called the Hill equation (or the Mathieu equation in the special case of an oscillating  perturbation). It is well known that there is an instability. In physics, the effect is known as {\bf parametric resonance} (see e.g. \cite{parres}). At frequencies $\omega_n$ corresponding to half integer multiples of the frequency $\omega$ of
the variation of the mass, i.e.
\be
\omega_k^2 = k_p^2 + m_{\chi}^2 \, = \, ({n \over 2} \omega)^2 \,\,\,\,\,\,\, n = 1, 2, ... ,
\ee
there are instability bands with widths $\Delta \omega_n$. For values of $\omega_k$ within the instability band, the value of $\chi_k$ increases exponentially:
\be
\chi_k \, \sim \, e^{\mu t} \,\,\,\, {\rm with} \,\,\, \mu \sim {{g^2 \sigma \varphi_0} \over {\omega}} \, ,
\ee
with $\varphi_0$ being the amplitude of the oscillation of $\varphi$. Since the widths of the instability bands decrease as a power of the (small) coupling constant $g^2$ with increasing $n$, for practical purposes only the lowest instability band is important. Its width is
\be
\Delta \omega_k \, \sim \, g \sigma^{1/2} \varphi_0^{1/2} \, .
\ee
Note, in particular, that there is no ultraviolet divergence in computing the total energy transfer from the $\varphi$ to the $\chi$ field due to parametric resonance.

It is easy to include the effects of the expansion of the Universe (see e.g. \cite{TB90,KLS94,STB95}). The main effect is that the value of $\omega_k$ becomes time-dependent. Thus, (for the model with a bare inflaton mass which we are considering) a mode slowly enters and leaves the resonance bands. As a consequence, any mode lies in the resonance band for only a finite time. This implies that the calculation of energy transfer is perfectly well-behaved. No infinite time divergences arise.{\footnote{Note, however, that even without expansion, scattering of the produced particles leads to a cutoff of the instability after some finite time (see e.g. \cite{Prok96} for a recent numerical study).}}

It is now possible to estimate the rate of energy transfer, whose order of magnitude is given by the phase space volume of the lowest instability band multiplied by the rate of growth of the mode function $\chi_k$. Using as an initial condition for $\chi_k$ the value $\chi_k \sim H$ given by the magnitude of the expected quantum fluctuations, we obtain
\be \label{entransf}
{\dot \rho} \, \sim \, \mu ({\omega \over 2})^2 \Delta\omega_k H e^{\mu t} \, .
\ee

From (\ref{entransf}) it follows that provided that the condition
\be \label{rescond}
\mu \Delta t \, >> 1
\ee
is satisfied, where $\Delta t < H^{-1}$ is the time a mode spends in the instability band, then the energy transfer will procede fast on the time scale
of the expansion of the Universe. In this case, there will be explosive particle production, and the energy density in matter at the end of reheating will be given by the energy density at the end of inflation.  

The above is a summary of the main physics of the modern theory of reheating.
The actual analysis can be refined in many ways (see e.g. \cite{KLS94,STB95}).
First of all, it is easy to take the expansion of the Universe into account
explicitly (by means of a transformation of variables), to employ an exact solution of the background model and to reduce the mode equation for $\chi_k$ to a Hill equation.

The next improvement consists of treating the $\chi$ field quantum mechanically (keeping $\varphi$ as a classical background field). At this point, the techniques of quantum field theory in a curved background can be applied. There is no need to impose artificial classical initial conditions for $\chi_k$. Instead, we may assume that $\chi$ starts in its initial vacuum state (excitation of an initial thermal state has been studied in \cite{Yoshimura2}), and the Bogoliubov mode mixing technique (see e.g. \cite{Birrell}) can be used to compute the number of particles at late times.

Using this improved analysis, we recover the result (\ref{entransf}). Thus, provided that the condition (\ref{rescond}) is satisfied, reheating will be explosive. Working out the time $\Delta t$ that a mode remains in the instability band for our model, expressing $H$ in terms of $\varphi_0$ and $m_{pl}$, and $\omega$ in terms of $\sigma$, and using the naturalness relation $g^2 \sim \lambda$, the condition for explosive particle production becomes
\be \label{rescond2}
{{\varphi_0 m_{pl}} \over {\sigma^2}} \, >> \, 1 \, ,
\ee
which is satisfied for all chaotic inflation models with $\sigma < m_{pl}$ (recall that slow rolling ends when $\varphi \sim m_{pl}$ and that therefore the initial amplitude $\varphi_0$ of oscillation is of the order $m_{pl}$).

We conclude that rather generically, reheating in chaotic inflation models will be explosive. This implies that the energy density after reheating will be approximately equal to the energy density at the end of the slow rolling period. Therefore, as suggested in \cite{KLS96,Tkachev} and \cite{KLR96}, respectively, GUT scale defects may be produced after reheating and GUT-scale baryogenesis scenarios may be realized, provided that the GUT energy scale is lower than
the energy scale at the end of slow rolling.

Note, however, that the state of $\chi$ after parametric resonance is {\bf not} a thermal state. The spectrum consists of high peaks in distinct wave bands. An important question which remains to be studied is how this state thermalizes.
For some interesting work on this issue see \cite{therm,Prok96}. As emphasized in \cite{KLS96} and \cite{Tkachev}, the large peaks in the spectrum may lead to symmetry restoration and to the efficient production of topological defects (for a differing view on this issue see \cite{AC96,Boyan2}). Since the state after explosive particle production is not a thermal state, it is useful to follow
\cite{KLS94} and call this process ``preheating" instead of reheating.

A futher interesting conjecture which emerges from the parametric resonance analysis of preheating$^{\cite{KLS94,STB95}}$ is that the dark matter in the Universe may consist of remnant coherent oscillations of the inflaton field. In fact, it can easily be checked from (\ref{rescond2}) that the condition for efficient transfer of energy eventually breaks down when $\varphi_0$ has decreased to a sufficiently small value. For the model considered here, an order of magnitude calculation shows that the remnant oscillations may well contribute significantly to the present value of $\Omega$.

Note that the details of the analysis of preheating are quite model-dependent. In fact$^{\cite{KLS94}}$, in many models one does not get the kind of ``narrow-band" resonance discussed here, but ``wide-band" resonance. In this case, the energy transfer is even more efficient.

There has recently been a lot of work on various aspects of reheating (see e.g. \cite{Yoshimura1,Boyan1,Kaiser,ALR96} for different approaches). Many important questions, e.g. concerning thermalization and back-reaction effects during and after preheating (or parametric resonance) remain to be fully analyzed.

\subsection*{Summary}

The inflationary Universe is an attractive scenario for early Universe cosmology. It can resolve some of the problems of standard cosmology, and in addition gives rise to a predictive theory of structure formation (see e.g. \cite{Liddle} for a recent review).

However, important unsolved problems of principle remain. Rather generically, the predicted amplitude of perturbations is too large (the spectral shape, however, is in quite good agreement with the observations). The present realizations of inflation based on scalar field also make the cosmological constant problem more accute. In addition, there are no convincing particle-physics based realizations of inflation. Many models of inflation resort to introducing a new matter sector. It is important to search for a better connection between modern particle physics / field theory and inflation.
String cosmology and dilaton gravity (see e.g. the recent reviews in \cite{stringcos}) may provide an interesting new approach to the unification of inflation and fundamental physics.

Recently, there has been much progress in the understanding of the energy transfer at the end of inflation between the inflaton field and matter. It appears that resonance phenomena such as parametric resonance play a crucial role. These new reheating scenarios lead to a high reheating temperature,
although much more work remains to be done before one can reach a final conclusion on this issue.

\section*{Topological Defects and Structure Formation}

\subsection*{Quantifying Data on Large-Scale Structure}

It is length scales corresponding to galaxies and larger which are of greatest
interest in cosmology when attempting to find an imprint of the primordial
fluctuations produced by particle physics.  On these scales, gravitational
effects are assumed to be dominant, and the fluctuations are not too far from
the linear regime.  On smaller scales, nonlinear gravitational and
hydrodynamical effects determine the final state and mask the initial
perturbations.

To set the scales, consider the mean separation of galaxies, which is about
5$h^{-1}$ Mpc$^{\cite{galaxy}}$, and that of Abell clusters which is around 25$h^{-1}$
Mpc$^{\cite{cluster}}$.  The largest coherent structures seen in current redshift surveys have a length of about 
100$h^{-1}$ Mpc$^{\cite{CFA,LCRS}}$, the recent detections of CMB
anisotropies probe the density field on length scales of about $10^3 h^{-1}$
Mpc, and the present horizon corresponds to a distance of about $3 \cdot 10^3
h^{-1}$ Mpc.

Galaxies are gravitationally bound systems containing billions of stars.  They
are non-randomly distributed in space.  A quantitative measure of this
non-randomness is the ``two-point correlation function" $\xi_2 (r)$ which gives
the excess probability of finding a galaxy at a distance $r$ from a given
galaxy:
\be
\xi_2 (r) = < \, {n (r) - n_0\over n_0} \, >  \, . %\eqno\eq
\ee
Here, $n_0$ is the average number density of galaxies, and $n(r)$ is the
density of galaxies a distance $r$ from a given one.  The pointed braces stand
for ensemble averaging.

Recent observational results from a various galaxy redshift surveys yield reasonable agreement$^{\cite{twopoint}}$ with a form 
\be
\xi_2 (r) \simeq \left({r_0\over r} \right)^\gamma %\eqno\eq
\ee
with scaling length $r_0 \simeq 5 h^{-1}$ Mpc and power $\gamma \simeq 1.8$.  A
theory of structure formation must explain both the amplitude and the slope of
this correlation function. 

On scales larger than galaxies, a better way to quantify structure is by means of large-scale systematic redshift surveys. Such surveys have
discovered coherent  planar structures and voids on scales of up
to $100 h^{-1}$ Mpc. Fig. 7 is a
sketch of redshift $z$ versus angle $\alpha$ in the sky for one $6^o$ slice of the sky$^{\cite{CFA}}$.  The second
direction in the sky has been projected onto the $\alpha -z$ plane.  The most
prominent feature is the band of galaxies at a distance of about $100h^{-1}$
Mpc.  This band also appears in neighboring slices and is therefore presumably
part of a planar density enhancement of comoving planar size of at least $(50
\times 100) \times h^{-2}$ Mpc$^2$.  This structure is often called the ``Great
Wall."  The challenge for theories of structure formation is not only to explain the fact that galaxies are nonrandomly distributed, but also to predict both
the observed scale and topology of the galaxy distribution. Topological defect models of structure formation attempt to address these questions.

Until 1992 there was little evidence for any convergence of the galaxy
distribution towards homogeneity.   Each new survey led to the discovery of new
coherent structures in the Universe on a scale comparable to that of the
survey.  In 1996, results of a much deeper redshift survey were
published$^{\cite{LCRS}}$ which for the first time find no coherent structures on the scale of the entire survey. In fact, no coherent structures on 
scales larger than $100 h^{-1}$ Mpc are seen.  This is the first direct evidence for the cosmological principle from optical surveys (the isotropy of the CMB has for a long time been a strong point in its support).

\begin{figure}
\centerline{\epsfig{file=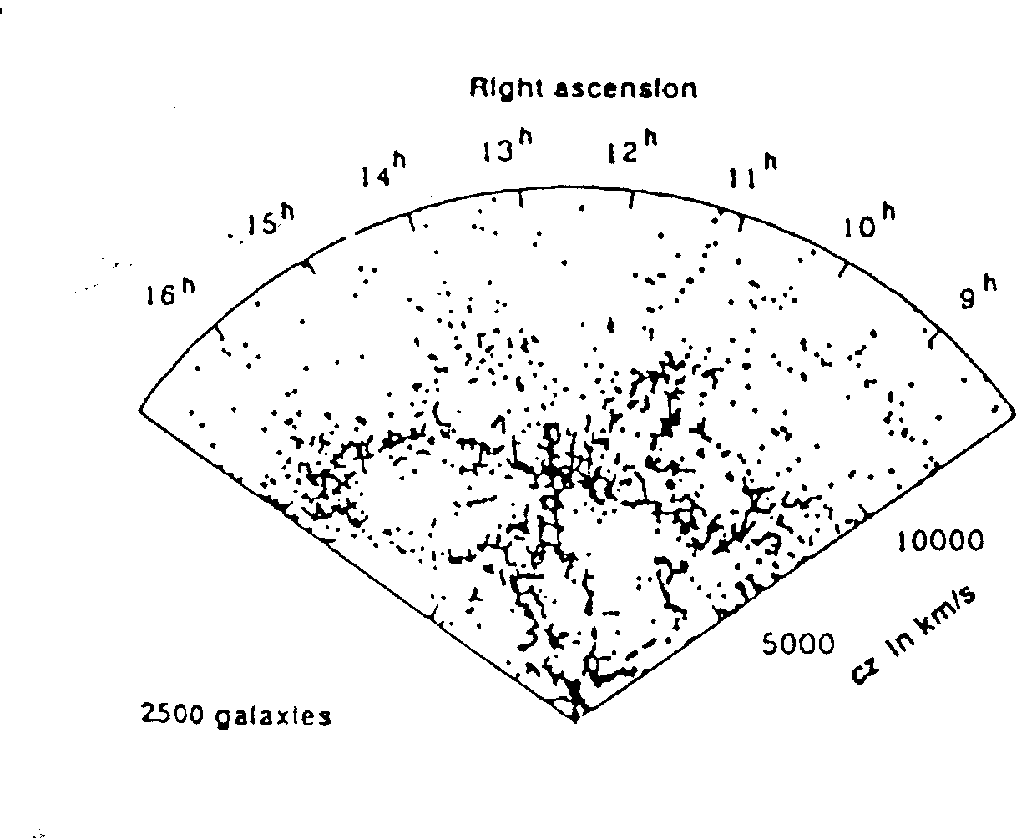,height=3in,width=3.5in}}
\caption{Results from the CFA redshift survey. Radial distance
gives the redshift of galaxies, the angular distance corresponds to right
ascension. The results from several slices of the sky (at different
declinations) have been projected into the same cone.}
\label{rhbfig7}
\end{figure}

In summary, a lot of data from optical and infrared galaxies are
currently available, and new data are being collected at a rapid rate.  The
observational constraints on theories of structure formation are becoming
tighter.   

\subsection*{Toplogical Defects}

According to particle physics theories, matter at high energies and
temperatures must be described in terms of fields.  Gauge symmetries have
proved to be extremely useful in describing the standard model of particle
physics, according to which at high energies the laws of nature are invariant
under a nonabelian group  $G$ of internal symmetry transformations
$G = {\rm SU} (3)_c \times {\rm SU} (2)_L \times U(1)_Y$
which at a temperature of about 200 MeV is spontaneously broken down to
$G^\prime = {\rm SU} (3)_c \times {\rm U} (1)$. The subscript on the SU(3) subgroup indicates that it is the color symmetry
group of the strong interactions, ${\rm SU} (2)_L \times $ U(1)$_Y$ is the
Glashow-Weinberg-Salam (WS) model of weak and electromagnetic interactions, the
subscripts $L$ and $Y$ denoting left handedness and hypercharge respectively.
At low energies, the WS model spontaneously breaks to the U(1) subgroup of
electromagnetism.

Spontaneous symmetry breaking is induced by an order parameter $\varphi$ taking
on a nontrivial expectation value $< \varphi >$ below a certain temperture
$T_c$.  In some particle physics models, $\varphi$ is a fundamental scalar
field in a nontrivial representation of the gauge group $G$ which is broken.
However, $\varphi$ could also be a fermion condensate, as in the BCS theory of superconductivity.

Earlier we have seen that symmetry breaking phase
transitions in gauge field theories do not, in general, lead to
inflation.  In most models, the coupling constants which arise in the
effective potential for the scalar field $\varphi$ driving the phase
transition are too large to generate a period of slow rolling which
lasts more than one Hubble time $H^{-1} (t)$.  Nevertheless, there are
interesting remnants of the phase transition: topological defects.
 
Consider a single component real scalar field with a typical symmetry breaking
potential
\be \label{stringpot}
V (\varphi) = {1\over 4} \lambda (\varphi^2 - \eta^2)^2 %\eqno\eq
\ee
Unless $\lambda \ll 1$ there
will be no inflation.  The phase transition will take place on a short time
scale $\tau < H^{-1}$, and will lead to correlation regions of radius $\xi <
t$ inside of which $\varphi$ is approximately constant, but outside of which
$\varphi$ ranges randomly over the vacuum manifold ${\cal M}$, the set of
values
of $\varphi$ which minimizes $V(\varphi)$ -- in our example $\varphi
= \pm \eta$.  The correlation regions are separated by domain walls, regions in
space where $\varphi$ leaves the vacuum manifold ${\cal M}$ and where,
therefore, potential energy is localized.  Via the usual gravitational
force, this energy density can act as a seed for structure.

Topological defects are familiar from solid state and condensed matter
systems.  Crystal defects, for example, form when water freezes or
when a metal crystallizes$^{\cite{Mermin}}$.  Point defects, line defects and planar
defects are possible.  Defects are also common in liquid crystals$^{\cite{LQ}}$.
They arise in a temperature quench from the disordered to the ordered
phase.  Vortices in $^4$He are analogs of global cosmic strings.
Vortices and other defects are also produced$^{\cite{Salomaa}}$ during a quench below the
critical temperature in $^3$He.  Finally, vortex lines may play an
important role in the theory of superconductivity$^{\cite{Abrikosov}}$.

The analogies between defects in particle physics and condensed matter
physics are quite deep.  Defects form for the same reason: the vacuum
manifold is topologically nontrivial.  The arguments$^{\cite{Kibble1}}$ which say that in
a theory which admits defects, such defects will inevitably form, are
applicable both in cosmology and in condensed matter physics.
Different, however, is the defect dynamics.  The motion of defects in
condensed matter systems is friction-dominated, whereas the defects in
cosmology obey relativistic equations, second order in time
derivatives, since they come from a relativistic field theory.

After these general comments we turn to a classification of
topological defects$^{\cite{Kibble1}}$.  We consider theories with an $n$-component order
parameter $\varphi$ and with a potential energy function (free energy
density) of the form (6.1) with $\varphi^2 = \sum\limits^n_{i = 1} \, \varphi^2_i$. 

There are various types of local and global topological defects
(regions of trapped energy density) depending on the number $n$ of components
of $\varphi$ (see e.g. \cite{VilShell} for a comprehensive survey of topological defect models).
The more rigorous mathematical definition refers to the homotopy
of ${\cal M}$.  The words ``local" and ``global" refer to whether the symmetry
which is broken is a gauge or global symmetry.  In the case of local
symmetries, the topological defects have a well defined core outside of which
$\varphi$ contains no energy density in spite of nonvanishing gradients
$\nabla \varphi$:  the gauge fields $A_\mu$ can absorb the gradient,
{\it i.e.,} $D_\mu \varphi = 0$ when $\partial_\mu \varphi \neq 0$,
where the covariant derivative $D_\mu$ is defined by
$D_\mu = \partial_\mu + ie \, A_\mu$, $e$ being the gauge coupling constant.
Global topological defects, however, have long range density fields and
forces.
\par
Table 1 contains a list of topological defects with their topological
characteristics.  A ``v" marks acceptable theories, a ``x" theories which are
in conflict with observations (for $\eta \sim 10^{16}$ GeV).

\begin{table}
\caption{Classification of cosmologically allowed (v) and forbidden (x) 
defects}
\label{table1}
%\begin{tabular}{lrrr}
%\begin{tabular}{lccc}
\begin{tabular}{lddd}
   defect name& n&
   \multicolumn{1}{c}{local defect} &
  \multicolumn{1}{c}{global defect}\\
\tableline
domain wall  & 1 & x & x \\
cosmic string & 2 & v & v \\
monopole & 3 & x & v \\
texture & 4 & - & v \\
\end{tabular}
\end{table}
   
Theories with domain walls are ruled out$^{\cite{nodw}}$ since a single domain wall stretching
across the Universe today would overclose the Universe.  Local monopoles are
also ruled out$^{\cite{nomon}}$ since they would overclose the Universe.  Local
textures are ineffective at producing structures because there is no traped potential energy.

From now on we will focus on one type of defects: cosmic strings (see e.g. \cite{VilShell,HK95,RB94} for recent reviews, and \cite{Vil85} for a classic review paper). These arise
in theories with a complex order parameter $(n = 2)$. In this case the vacuum manifold of the model is
\be
{\cal M} = S^1 \, , %\eqno\eq
\ee
which has nonvanishing first homotopy group:
\be
\Pi_1 ({\cal M}) = Z \neq 1 \, . %\eqno\eq
\ee
A cosmic string is a line of trapped energy density which arises
whenever the field $\varphi (x)$ circles ${\cal M}$ along a closed path
in space ({\it e.g.}, along a circle).  In this case, continuity of
$\varphi$ implies that there must be a point with $\varphi = 0$ on any
disk whose boundary is the closed path.  The points on different sheets
connect up to form a line overdensity of field energy.  

To construct a field configuration with a string along the $z$ axis$^{\cite{Nielsen}}$,
take $\varphi (x)$ to cover ${\cal M}$ along a circle with radius $r$
about the point $(x,y) = (0,0)$:
\be \label{stringconf}
\varphi (r, \vartheta ) \simeq \eta e^{i \vartheta} \, , \, r \gg
\eta^{-1} \, . %\eqno\eq
\ee
This configuration has winding number 1, {\it i.e.}, it covers ${\cal
M}$ exactly once.  Maintaining cylindrical symmetry, we can extend
(\ref{stringconf}) to arbitrary $r$
\be
\varphi (r, \, \vartheta) = f (r) e^{i \vartheta} \, , %\eqno\eq
\ee
where $f (0) = 0$ and $f (r)$ tends to $\eta$ for large $r$.  The
width $w$ can be found by balancing potential and tension
energy.  The result is 
\be
w \, \sim \, \lambda^{-1/2} \eta^{-1} \, .
\ee

For local cosmic strings, {\it i.e.}, strings arising due to the
spontaneous breaking of a gauge symmetry, the energy density decays
exponentially for $r \gg w$.  In this case, the energy $\mu$
per unit length of a string is finite and depends only on the symmetry
breaking scale $\eta$
\be
\mu \sim \eta^2 %\eqno\eq
\ee
(independent of the coupling $\lambda$).  The value of $\mu$ is the
only free parameter in a cosmic string model.

\subsection*{Formation of Defects in Cosmological Phase Transitions}

The symmetry breaking phase transition takes place at $T = T_c$ (called the critical temperature).  From condensed matter physics it is well
known that in many cases topological defects form during phase transitions,
particularly if the transition rate is fast on a scale compared to the system
size.  When cooling a metal, defects in the crystal configuration will be
frozen in; during a temperature quench of $^4$He, thin vortex tubes of the
normal phase are trapped in the superfluid; and analogously in a temperature
quench of a superconductor, flux lines are trapped in a surrounding sea of the
superconducting Meissner phase.

In cosmology, the rate at which the phase transition proceeds is given by the
expansion rate of the Universe.  Hence, topological defects will inevitably be
produced in a cosmological phase transition$^{\cite{Kibble1}}$, provided the underlying particle physics model allows such defects. 
 
\begin{figure}[b!]
\centerline{\epsfig{file=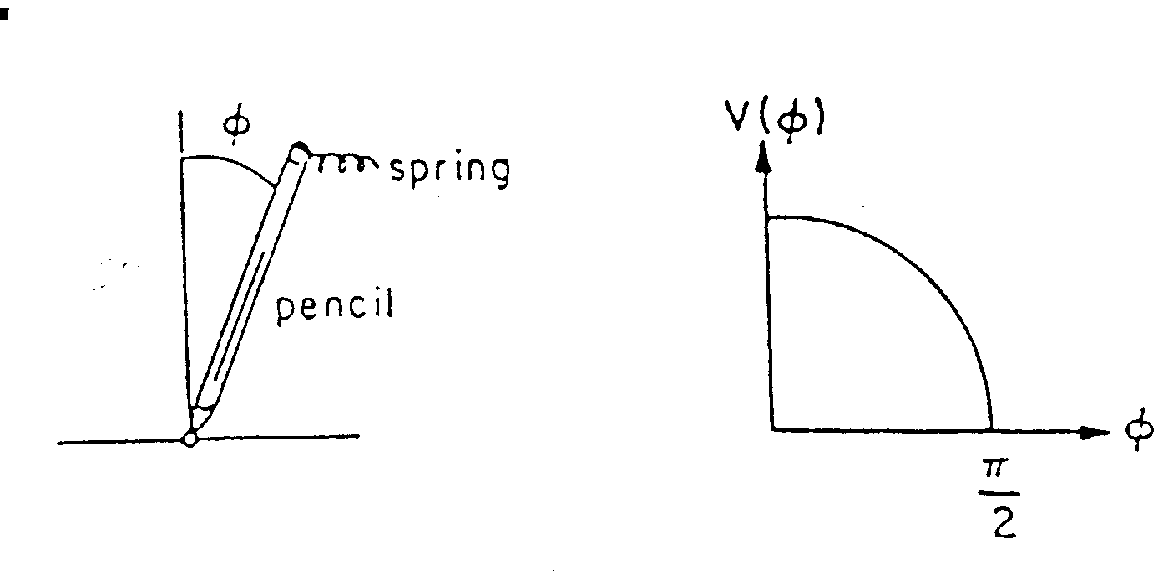,height=2.5in,width=3.5in}}
\caption{The pencil model: the potential energy of a
simple pencil has the same form as that of scalar fields used for
spontaneous symmetry breaking.  The springs connecting nearest
neighbor pencils give rise to contributions to the energy which mimic
spatial gradient terms in field theory.}
\label{rhbfig8}
\end{figure}

The argument which  ensures that in theories which admit
topological or semitopological defects, such defects will be produced
during a phase transition in the very early Universe is called the Kibble mechanism$^{\cite{Kibble1}}$. To illustrate the physics, consider
a mechanical toy model, first introduced by Mazenko, Unruh
and Wald$^{\cite{MUW}}$. Take (see Fig.
\ref{rhbfig8}) a lattice of points on a flat table.  At each point, a pencil
is pivoted.  It is free to rotate and oscillate.  The tips of nearest
neighbor pencils are connected with springs (to mimic the spatial
gradient terms in the scalar field Lagrangean).  Newtonian gravity
creates a potential energy $V(\varphi)$ for each pencil ($\varphi$ is
the angle relative to the vertical direction).  $V(\varphi)$ is
minimized for $| \varphi | = \eta$ (in our toy model $\eta = \pi /
2$).  Hence, the Lagrangean of this pencil model is analogous to that
of a scalar field with symmetry breaking potential (\ref{stringpot}).

At high temperatures $T \gg T_c$, all pencils undergo large amplitude
high frequency oscillations.  However, by causality, the phases of
oscillation of pencils with large separation $s$ are uncorrelated.
For a system in thermal equilibrium, the length $s$ beyond which
phases are random is the correlation length $\xi (t)$.  However, since
the system is quenched rapidly, there is a causality bound on
$\xi$:
\be
\xi (t) < t \, , %\eqno\eq
\ee
where $t$ is the causal horizon.

The critical temperature $T_c$ is the temperature at which the
thermal energy is equal to the energy a pencil needs to jump from
horizontal to vertical position.  For $T < T_c$, all pencils want to
lie flat on the table.  However, their orientations are random beyond
a distance $\xi (t)$ determined by equating the free energy gained by
symmetry breaking (a volume effect) with the gradient energy lost (a surface
effect).  As expected, $\xi (T)$ diverges at $T_c$. Very close to $T_c$, the
thermal energy $T$ is larger than the volume energy gain $E_{corr}$ in a
correlation volume. Hence, these domains are unstable to thermal fluctuations.
As $T$ decreases, the thermal energy decreases more rapidly than $E_{corr}$.
Below the ``Ginsburg temperature" $T_G$, there
is insufficient thermal energy to excite a correlation volume into the
state $\varphi = 0$.  Domains of size
\be \label{corrlength}
\xi (t_G) \sim \lambda^{-1} \eta^{-1} %\eqno\eq
\ee
freeze out$^{\cite{Kibble1,Kibble2}}$.  The boundaries between these domains become
topological defects. An improved version of this argument has recently been given by Zurek$^{\cite{Zurek2}}$ (see also \cite{BDH}).

We conclude that in a theory in which a symmetry breaking phase
transitions satisfies the topological criteria for the existence of a
fixed type of defect, a network of such defects will form during the
phase transition and will freeze out at the Ginsburg temperature.  The
correlation length is initially given by (\ref{corrlength}), if the field
$\varphi$ is in thermal equilibrium before the transition.
Independent of this last assumption, the causality bound implies that
$\xi (t_G) < t_G$.

For times $t > t_G$ the evolution of the network of defects may be
complicated (as for cosmic strings) or trivial (as for textures).  In
any case (see the caveats of \cite{caveat}), the causality bound
persists at late times and states that even at late times, the mean
separation and length scale of defects is bounded by $\xi (t) \leq t$.

Applied to cosmic strings, the Kibble mechanism implies that at the
time of the phase transition, a network of cosmic strings with typical
step length $\xi (t_G)$ will form.  According to numerical
simulations$^{\cite{VachVil}}$, about 80\% of the initial energy is in infinite
strings (strings with curvature radius larger than the Hubble radius) and 20\% in closed loops.

\subsection*{Evolution of Strings and Scaling}

\begin{figure}[b!]
\centerline{\epsfig{file=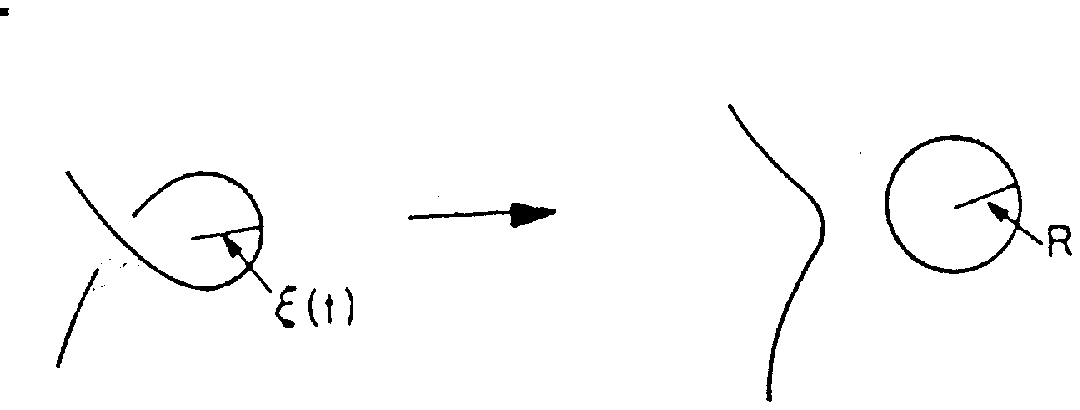,height=1.5in,width=3.5in}}
\caption{Formation of a loop by a self intersection of an
infinite string. According to the original cosmic string scenario, loops form
with a radius $R$ determined by the instantaneous coherence length of the
infinite string network.}
\label{rhbfig9}
\end{figure}

The evolution of the cosmic string network for $t > t_G$ is
complicated.  The key processes are loop production
by intersections of infinite strings (see Fig. \ref{rhbfig9}) and loop shrinking
by gravitational radiation.  These two processes combine to create a
mechanism by which the infinite string network loses energy (and
length as measured in comoving coordinates).  It can be shown (see e.g. \cite{Vil85}) that, as
a consequence, the correlation length of the string network is always
proportional to its causality limit
\be
\xi (t) \sim t \, . %\eqno\eq
\ee
Hence, the energy density $\rho_\infty (t)$ in long strings is a fixed
fraction of the background energy density $\rho_c (t)$
\be
\rho_\infty (t) \sim \mu \xi (t)^{-2} \sim \mu t^{-2} %\eqno\eq
\ee
or
\be
{\rho_\infty (t)\over{\rho_c (t)}} \sim G \mu \, . %\eqno\eq
\ee

We conclude that the cosmic string network approaches a ``scaling
solution" in which the statistical properties of the
network are time independent if all distances are scaled to the
horizon distance.
  
\subsection*{Cosmic Strings and Structure Formation}

\begin{figure}[b!]
\centerline{\epsfig{file=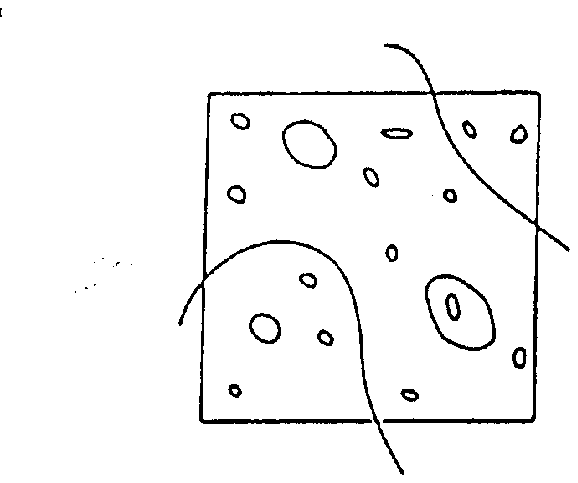,height=2.5in,width=3.5in}}
\caption{ Sketch of the scaling solution for the
cosmic string network.  The box corresponds to one Hubble volume at
arbitrary time $t$.}
\label{rhbfig10}
\end{figure}

The starting point of the structure formation scenario in the cosmic
string theory is the scaling solution for the cosmic string network,
according to which at all times $t$ (in particular at $t_{eq}$, the
time when perturbations can start to grow) there will be a few long
strings crossing each Hubble volume, plus a distribution of loops of
radius $R \ll t$ (see Fig. \ref{rhbfig10}).

The cosmic string model admits three mechanisms for structure
formation:  loops, filaments, and wakes.  Cosmic string loops have the same
time averaged field as a point source with mass$^{\cite{Turok84}}$
$ M (R) = \beta R \mu $, $R$ being the loop radius and $\beta \sim 2 \pi$.  Hence, loops will be seeds for spherical accretion of dust and radiation.

For loops with $R \leq t_{eq}$, growth of perturbations in a model
dominated by cold dark matter starts at $t_{eq}$.  Hence, the mass at
the present time will be
\be
M (R, \, t_0) = z (t_{eq}) \beta \, R \mu \, . %\eqno\eq
\ee

In the original cosmic string model$^{\cite{ZelVil,TB86}}$ it was assumed
that loops dominate over wakes.  However, according to the newer cosmic string evolution simulations$^{\cite{CSsimuls}}$, most of the energy in strings is in the long strings, and hence the loop accretion mechanism is subdominant. 

\begin{figure}[b!]
\centerline{\epsfig{file=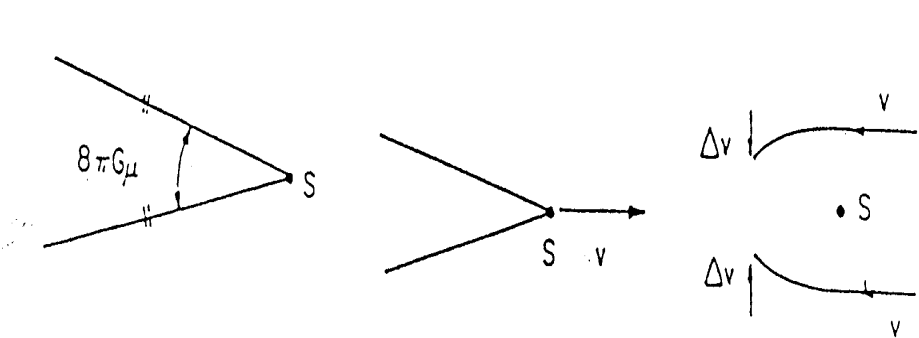,height=2in,width=5.5in}}
\caption{Sketch of the mechanism by which a long
straight cosmic string $S$ moving with velocity $v$ in transverse
direction through a plasma induces a velocity perturbation $\Delta v$
towards the wake. Shown on the left is the deficit angle, in the
center is a sketch of the string moving in the plasma, and on the
right is the sketch of how the plasma moves in the frame in which the
string is at rest.}
\label{rhbfig11}
\end{figure}

The second mechanism involves long strings moving with relativistic
speed in their normal plane which give rise to
velocity perturbations in their wake$^{\cite{SilkVil}}$.  The mechanism is illustrated in Fig. \ref{rhbfig11}:
space normal to the string is a cone with deficit angle$^{\cite{Vil81}}$
\be \label{deficit}
\alpha = 8 \pi G \mu \, . 
\ee
If the string is moving with normal velocity $v$ through a bath of dark
matter, a velocity perturbation
\be
\delta v = 4 \pi G \mu v \gamma  
\ee
[with $\gamma = (1 - v^2)^{-1/2}$] towards the plane behind the string
results.  At times after $t_{eq}$, this induces planar overdensities,
the most
prominent ({\it i.e.}, thickest at the present time) and numerous of which were
created at $t_{eq}$, the time of equal matter and
radiation$^{\cite{TV86,AS87,LP90}}$.  The
corresponding planar dimensions are (in comoving coordinates)
\be
t_{eq} z (t_{eq}) \times t_{eq} z (t_{eq}) v \sim (40 \times 40 v) \,
{\rm Mpc}^2\, . 
\ee

The thickness $d$ of these wakes can be calculated using the
Zel'dovich approximation$^{\cite{Zeld70}}$.  The result is (for $G \mu = 10^{-6}$)
\be
d \simeq G \mu v \gamma (v) z (t_{eq})^2 \, t_{eq} \simeq 4 v \, {\rm
Mpc} \, . 
\ee
 
Wakes arise if there is little small scale structure on the string.
In this case, the string tension equals the mass density, the string
moves at relativistic speeds, and there is no local gravitational
attraction towards the string.

In contrast, if there is small scale structure on strings,
then the coarse-grained string tension $T$ is smaller$^{\cite{Carter}}$ 
than the mass per unit length $\mu$ , and thus there
is a gravitational force towards the string which gives rise to
cylindrical accretion, producing filaments$^{\cite{fils}}$.

Which of the mechanisms -- filaments or wakes -- dominates is
determined by the competition between the velocity induced by the Newtonian gravitational potential of the strings and the velocity perturbation of the wake.   

The cosmic string model predicts a scale-invariant spectrum of density
perturbations, exactly like inflationary Universe models but for a
rather different reason.  Consider the {\it r.m.s.} mass fluctuations
on a length scale $2 \pi k^{-1}$ at the time $t_H (k)$ when this scale
enters the Hubble radius.  From the cosmic string scaling solution it
follows that a fixed ({\it i.e.}, $t_H (k)$ independent) number
$\tilde v$ of strings of length of the order $t_H (k)$ contribute to
the mass excess $\delta M (k, \, t_H (k))$.  Thus
\be
{\delta M\over M} \, (k, \, t_H (k)) \sim \, {\tilde v \mu t_H
(k)\over{G^{-1} t^{-2}_H (k) t^3_H (k)}} \sim \tilde v \, G \mu \, .
\ee
Note that the above argument predicting a scale invariant spectrum
will hold for all topological defect models which have a scaling
solution, in particular also for global monopoles and textures.

The amplitude of the {\it r.m.s.} mass fluctuations (equivalently: of
the power spectrum) can be used to normalize $G \mu$.  Since today on
galaxy cluster scales
\be
{\delta M\over M} (k, \, t_0) \sim 1 \, , %\eqno\eq
\ee
the growth rate of fluctuations linear in $a(t)$ yields
\be
{\delta M\over M} \, (k, \, t_{eq}) \sim 10^{-4} \, , %\eqno\eq
\ee
and therefore, using $\tilde v \sim 10$,
\be
G \mu \sim 10^{-5} \, . %\eqno\eq
\ee
A similar value is obtained by normalizing the model to the COBE amplitude of CMB anisotropies on large angular scales$^{\cite{BBS,Periv}}$ (the normalizations from COBE and from the power spectrum of density perturbations on large scales agree to within a factor of 2).
Thus, if cosmic strings are to be relevant for structure formation,
they must arise due to a symmetry breaking at energy scale $\eta
\simeq 10^{16}$GeV.  This scale happens to be the scale of unification (GUT)
of weak, strong and electromagnetic interactions.  It is tantalizing
to speculate that cosmology is telling us that there indeed was new
physics at the GUT scale.

A big advantage of the cosmic string model over inflationary Universe
models is that HDM is a viable dark matter candidate.  Cosmic string
loops survive free streaming, as discussed in Section 3.B, and can
generate nonlinear structures on galactic scales, as discussed in
detail in \cite{BKST,Edbert}.  Accretion of hot dark matter by a string wake
was studied in \cite{LP90}. In this case, nonlinear perturbations
develop only late.  At some time $t_{nl}$, all scales up to a distance
$q_{\rm max}$ from the wake center go nonlinear.  Here
\be
q_{\rm max} \sim G \mu v \gamma (v) z (t_{eq})^2 t_{eq} \sim 4 v \,
{\rm Mpc} \, , %\eqno\eq
\ee
and it is the comoving thickness of the wake at $t_{nl}$.  Demanding
that $t_{nl}$ corresponds to a redshift greater than 1 leads to the
constraint
\be
G \mu > 5 \cdot 10^{-7} \, . %\eqno\eq
\ee
Note that in a cosmic string and hot dark matter model, wakes form nonlinear structures only very recently. Accretion onto loops and small scale structure on the long strings provide two mechanisms which may lead to high redshift objects such as quasars and high redshift galaxies. The first mechanism has recently been studied in \cite{MB96}, the second in \cite{AB95,ZLB96}.

The power spectrum of density fluctuations in a cosmic string model with HDM has recently been studied numerically by M\"ah\"onen$^{\cite{Mahonen}}$, based on previous work of \cite{Hara} (see also \cite{AS92} for an earlier semi-analytical study). The spectral shape agrees quite well with observations, and a bias factor of less than 2 is required to give the best-fit amplitude for a COBE normalized model. Note, however, that the results depend quite sensitively on the details of the string scaling solution which are at present not well understood.

Due to lack of space, I will not discuss the global monopole$^{\cite{BenRhie}}$ and global texture$^{\cite{Turok89}}$ models of structure formation. There has been a lot of work on the texture model, and the reader is referred to \cite{Turok91,Durrer94} for recent review articles.

\subsection*{Specific Signatures}

The cosmic string theory of structure formation makes several distinctive predictions, both in terms of the galaxy distribution and in terms of CMB anisotropies. On large scales (corresponding to the comoving Hubble radius at $t_{eq}$ and larger, structure is predicted to be dominated either by planar$^{\cite{TV86,AS87,LP90}}$ or filamentary$^{\cite{fils}}$ galaxy concentrations. For models in which the strings have no local gravity, the resulting nonlinear structures will look very different from the nonlinear structures in models in which local gravity is the dominant force. As discovered and discussed recently in \cite{SB96}, a baryon number excess is predicted in the nonlinear wakes. This may explain the ``cluster baryon crisis"$^{\cite{clusterbaryon}}$, the fact that the ratio of baryons to dark matter in rich clusters is larger than what is compatible with the nucleosynthesis constraints in a spatially flat Universe.

\begin{figure}[b!]
\centerline{\epsfig{file=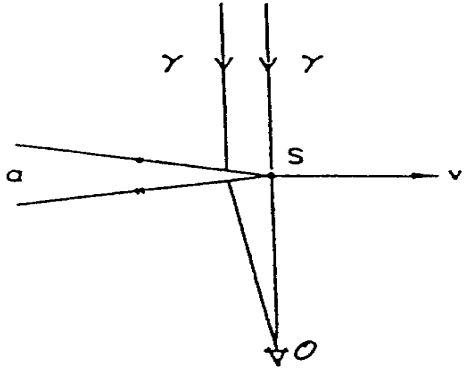,height=2.5in,width=3.5in}}
\caption{Sketch of the Kaiser-Stebbins effect by
which cosmic strings produce linear discontinuities in the CMB. Photons
$\gamma$ passing on different sides of a moving string $S$ (velocity $v$)
towards the observer ${\cal O}$ receive a relative Doppler shift due to the
conical nature of space perpendicular to the string (deficit angle $\alpha$).}
\label{rhbfig12}
\end{figure} 

As described in the previous subsection, space perpendicular to a long straight
cosmic string is conical with deficit angle given by (\ref{deficit}).  Consider
now CMB radiation approaching an observer in a direction normal to the
plane spanned by the string and its velocity vector (see Fig. \ref{rhbfig12}).
Photons arriving at the observer having passed on different sides of
the string will obtain a relative Doppler shift which translates into
a temperature discontinuity of amplitude$^{\cite{KS84}}$
\be
{\delta T\over T} = 4 \pi G \mu v \gamma (v) \, , %\eqno\eq
\ee
where $v$ is the velocity of the string.  Thus, the distinctive
signature for cosmic strings in the microwave sky are line
discontinuities in $T$ of the above magnitude.

Given ideal maps of the CMB sky it would be easy to detect strings.
However, real experiments have finite beam width.  Taking into account
averaging over a scale corresponding to the beam width will smear out
the discontinuities, and it turns out to be surprisingly hard to
distinguish the predictions of the cosmic string model from that of
inflation-based theories using quantitative statistics which are easy
to evaluate analytically, such as the kurtosis of the spatial gradient
map of the CMB$^{\cite{MPB94}}$. There may be ways to distinguish between string and inflationary models by looking at the angular power spectrum of CMB anisotropies. Work on this subject, however, is still controversial$^{\cite{AAcrew,HuWhite,Turok96}}$.

Global textures also produce distinctive non-Gaussian signatures$^{\cite{TurSper}}$ in CMB maps. In fact, these signatures are more pronounced and on larger scales than the signatures in the cosmic string model. 

\section*{Topological Defects and Baryogenesis}

\subsection*{Principles of Baryogenesis}

Baryogenesis is another area where particle physics and cosmology connect in a very deep way. The goal is to explain the observed asymmetry between matter and antimatter in the Universe. In particular, the objective is to be able to explain the observed value of the net baryon to entropy ratio at the present time
\be
{{\Delta n_B} \over s}(t_0) \, \sim \, 10^{-10} 
\ee
starting from initial conditions in the very early Universe when this ratio vanishes. Here, $\Delta n_B$ is the net baryon number density and $s$ the entropy density.

As pointed out by Sakharov$^{\cite{Sakharov}}$, three basic criteria must be satisfied in order to have a chance at explaining the data:
\begin{enumerate}
\item{} The theory describing the microphysics must contain baryon number violating processes.
\item{} These processes must be C and CP violating.
\item{} The baryon number violating processes must occur out of thermal equilibrium.
\end{enumerate}

As was discovered in the 1970's$^{\cite{GUTBG}}$, all three criteria can be satisfied in GUT theories. In these models, baryon number violating processes are mediated by superheavy Higgs and gauge particles. The baryon number violation is visible in the Lagrangian, and occurs in perturbation theory (and is therefore in principle easy to calculate). In addition to standard model CP violation, there are typically many new sources of CP violation in the GUT sector. The third Sakharov condition can also be realized: After the GUT symmetry-breaking phase transition, the superheavy particles may fall out of thermal equilibrium. The out-of-equilibrium decay of these particles can thus generate a nonvanishing baryon to entropy ratio. 

The magnitude of the predicted $n_B / s$ depends on the asymmetry $\varepsilon$ per decay, on the coupling constant $\lambda$ of the $n_B$ violating processes, and on the ratio $n_X / s$ of the number density $n_X$ of superheavy Higgs and gauge particles to the number density of photons, evaluated at the time $t_d$ when the baryon number violating processes fall out of thermal equilibrium, and assuming
that this time occurs after the phase transition. The quantity $\varepsilon$ is proportional to the CP-violation parameter in the model. In a GUT theory, this CP violation parameter can be large (order 1), whereas in the standard electroweak theory it is given by the CP violating phases in the CKM mass matrix and is very small. As shown in \cite{GUTBG} it is easily possible to construct models which give the right $n_B / s$ ratio after the GUT phase transition (for recent reviews of baryogenesis see \cite{Dolgov} and \cite{RubShap}).
 
\subsection*{GUT Baryogenesis and Topological Defects}

The ratio $n_B / s$, however, does not only depend on $\varepsilon$, but also on $n_X / s (t_d)$. If the temperature $T_d$ at the time $t_d$ is greater than the mass $m_X$ of the superheavy particles, then it follows from the thermal history in standard cosmology that $n_X \sim s$. However, if $T_d < m_X$, then the number density of $X$ particles is diluted exponentially in the time interval between when $T = m_X$ and when $T = T_d$. Thus, the predicted baryon to entropy ratio is exponentially suppressed:
\be \label{expdecay}
{n_B \over s} \, \sim \, {1 \over {g^*}} \lambda^2 \varepsilon e^{- m_X / T_d} \, ,
\ee
where $g^*$ is the number of spin degrees of freedom in thermal equilibrium at the time of the phase transition.
In this case, the standard GUT baryogenesis mechanism is ineffective.

However, topological defects may come to the rescue$^{\cite{BDH92}}$. As we have seen in the previous section, topological defects will inevitably be produced in the symmetry breaking GUT transition provided they are topologically allowed in that symmetry breaking scheme. The topological defects provide an alternative mechanism of GUT baryogenesis.

Inside of topological defects, the GUT symmetry is restored. In fact, the defects can be viewed as solitonic configurations of $X$ particles. The continuous decay of defects at times after $t_d$ provides an alternative way to generate a nonvanishing baryon to entropy ratio. The defects constitute out of equilibrium configurations, and hence their decay can produce a nonvanishing $n_B / s$ in the same way as the decay of free $X$ quanta. 

The way to compute the estimate $n_B / s$ ratio is as follows: The defect
scaling solution gives the energy density in defects at all times. Taking the time derivative of this density, and taking into account the expansion of the Universe, we obtain the loss of energy attributed to defect decay. By energetics, we can estimate the number of decays of individual quanta which the defect decay corresponds to. We can then use the usual perturbative results to compute the resulting net baryon number.

Provided that $m_X < T_d$, then at the time when the baryon number violating processes fall out of equilibrium (when we start generating a nonvanishing $n_B$) the energy density in free $X$ quanta is much larger than the defect density, and hence the defect-driven baryogenesis mechanism is subdominant. However, if $m_X > T_d$, then as indicated in (\ref{expdecay}), the energy density in free quanta decays exponentially. In constrast, the density in defects only
decreases as a power of time, and hence soon dominates baryogenesis.

One of the most important ingredients in the calculation is the time dependence of $\xi(t)$, the separation between defects. Immediately after the phase transition at the time $t_f$ of the formation of the defect network, the separation is $\xi(t_f) \sim \lambda^{-1} \eta^{-1}$. In the time period immediately following, the time period of relevance for baryogenesis, $\xi(t)$ approaches the Hubble radius according to the equation$^{\cite{Kibble2}}$
\be \label{defsep}
\xi(t) \, \simeq \, \xi(t_f) ({t \over {t_f}})^{5/4} \, .
\ee
Using this result to calculate the defect density, we obtain after some algebra
\be \label{barres}
{{n_B} \over s}|_{\rm defect} \, \sim \, \lambda^2 {{T_d} \over \eta} {{n_B} \over s}|_0 \, ,
\ee
where $n_B / s|_0$ is the unsuppressed value of $n_B / s$ which can be obtained using the standard GUT baryogenesis mechanism. We see from (\ref{barres}) that even for low values of $T_d$, the magnitude of $n_B / s$ which is obtained via the defect mechanism is only suppressed by a power of $T_d$. However, the maximum strength of the defect channel is smaller than the maximum strength of the usual mechanism by a geometrical suppression factor $\lambda^2$ which expresses the fact that even at the time of defect formation, the defect network only occupies a small volume.

\subsection*{Electroweak Baryogenesis and Topological Defects}

It has been known for some time that there are baryon number violating processes even in the standard electroweak theory. These processes are, however, nonperturbative. They are connected with the t'Hooft anomaly$^{\cite{tHooft}}$, which in turn is due to the fact that the gauge theory vacuum is degenerate, and that the different degenerate vacuum states have different quantum numbers (Chern-Simons numbers). In theories with fermions, this implies different baryon number. Configurations such as sphalerons$^{\cite{sphal}}$ which interpolate between two such vacuum states thus correspond to baryon number violating processes.

As pointed out in \cite{KRS85}, the anomalous baryon number violating processes are in thermal equilibrium above the electroweak symmetry breaking scale. Therefore, any net baryon to entropy ratio generated at a higher scale will be erased, unless this ratio is protected by an additional quantum number such as a nonvanishing $B - L$ which is conserved by electroweak processes.

However, as first suggested in \cite{Shap} and discussed in detail in many recent papers (see \cite{EWBGrev} for reviews of the literature), it is possible to regenerate a nonvanishing $n_B / s$ below the electroweak symmetry breaking scale. Since there are $n_B$ violating processes and both C and CP violation in the standard model, Sakharov's conditions are satisfied provided that one can realize an out-of-equilibrium state after the phase transition. Standard model CP violation is extremely weak. Thus, it appears necessary to add some sector with extra CP violation to the standard model in order to obtain an appreciable $n_B / s$ ratio. A simple possibility which has been invoked often is to add a second Higgs doublet to the theory, with CP violating relative phases. 

The standard way to obtain out-of-equilibrium baryon number violating processes immediately after the electroweak phase transition is$^{\cite{EWBGrev}}$ to assume that the transition is strongly first order and proceeds by the nucleation of bubbles (note that these are two assumptions, the second being stronger than the first!). 

Bubbles are out-of-equilibrium configurations. Outside of the bubble (in the false vacuum), the baryon number violating processes are unsuppressed, inside they are exponentially suppressed. In the bubble wall, the Higgs fields have a nontrivial profile, and hence (in models with additional CP violation in the Higgs sector) there is enhanced CP violation in the bubble wall. In order to obtain net baryon production, one may either use fermion scattering off bubble walls$^{\cite{CKN1}}$ (because of the CP violation in the scattering, this generates a lepton asymmetry outside the bubble which converts via sphalerons to a baryon asymmetry) or sphaleron processes in the bubble wall itself$^{\cite{TZ,CKN2}}$. It has been shown that, using optimistic parameters (in particular a large CP violating phase $\Delta \theta_{CP}$ in the Higgs sector) it is possible to generate the observed $n_B / s$ ratio. The resulting baryon to entropy ratio is of the order
\be \label{ewres}
{{n_B} \over s} \, \sim \, \alpha_W^2 (g^*)^{-1} \bigl( {{m_t} \over T} \bigr)^2 \Delta \theta_{CP} \, ,
\ee
where $\alpha_W$ refers to the electroweak interaction strength, $g^*$ is the number of spin degrees of freedom in thermal quilibrium at the time of the phase transition, and $m_t$ is the top quark mass. The dependence on the top quark mass enters because net baryogenesis only appears at the one-loop level.

However, analytical and numerical studies show that, for the large Higgs masses which are indicated by the current experimental bounds, the electroweak phase transition will unlikely be sufficiently strongly first order to proceed by bubble nucleation. In addition, there are some concerns as to whether it will proceed by bubble nucleation at all (see e.g. \cite{Gleiser}).

Once again, topological defects come to the rescue. In models which admit defects, such defects will inevitably be produced in a phase transition independent of its order. Moving topological defects can play the same
role in baryogenesis as nucleating bubbles. In the defect core, the electroweak symmetry is unbroken and hence sphaleron processes are unsuppressed$^{\cite{Perkins}}$. In the defect walls there is enhanced CP violation for the same reason as in bubble walls. Hence, at a fixed point in space, a nonvanishing baryon number will be produced when a topological defect passes by.

Defect-mediated electroweak baryogenesis has been worked out in detail in \cite{BDPT} (see \cite{BDT} for previous work) in the case of cosmic strings. The scenario is as follows: at a particular point $x$ in space, antibaryons are produced when the front side of the defect passes by. While $x$ is in the defect core, partial equilibration of $n_B$ takes place via sphaleron processes. As the back side of the defect passes by, the same number of baryons are produced as the number of antibaryons when the front side of the defect passes by. Thus, at the end a positive number of baryons are left behind.

As in the case of defect-mediated GUT baryogenesis, the strength of defect-mediated electroweak baryogenesis is suppressed by the ratio ${\rm SF}$ of the volume which is passed by defects divided by the total volume, i.e.
\be
{{n_B} \over s} \, \sim \, {\rm SF} {{n_B} \over s}|_0 \, ,
\ee
where $(n_B / s)|_0$ is the result of (\ref{ewres}) obtained in the bubble nucleation mechanism. 

A big caveat for defect-mediated electroweak baryogenesis is that the standard electroweak theory does not admit topological defects. However, in a theory with additional physics just above the electroweak scale it is possible to obtain defects (see e.g. \cite{TDB95} for some specific models). The closer the scale $\eta$ of the new physics is to the electroweak scale $\eta_{EW}$, the larger the volume in defects and the more efficient defect-mediated electroweak baryogenesis. Using the result of (\ref{defsep}) for the separation of defects, we obtain (for non-superconducting strings)
\be
{\rm SF} \, \sim \, \lambda \bigl( {{\eta_{EW}} \over \eta} \bigr)^{3/2} v_D\, .
\ee 
where $v_D$ is the mean defect velocity.
 
Obviously, the advantage of the defect-mediated baryongenesis scenario is that it does not depend on the order and on the detailed dynamics of the electroweak phase transition.  

\subsection*{Summary}

As we have seen, topological defects may play an important role in cosmology.
Defects are inevitably produced during symmetry breaking phase transitions in the early Universe in all theories in which defects are topologically stable.
Theories giving rise to domain walls or local monopoles are ruled out by cosmological constraints. Those producing cosmic strings, global monopoles and textures are quite attractive.

If the scale of symmetry breaking at which the defects are produced is about $10^{16}$ GeV, then defects can act as the seeds for galaxy formation. Defect theories of structure formation predict a roughly scale-invariant spectrum of density perturbations, similar to inflation-based models. However, the phases in the density field are distributed in a non-Gaussian manner. Thus, the predictions of defect models can be distinguished from those of inflationary models. In addition, the predictions of different defect models can be distinguished from eachother.

As shown in this section, topological defects may also play a crucial role in baryogenesis. This applies
both to GUT and electroweak baryogenesis. The crucial point is that defects constitute out-of-equilibrium configurations, and may therefore be the sites of net baryon production.

\medskip
\centerline{\bf Acknowledgements}
\medskip

I wish to thank the organizers of the school and symposium for inviting me to
speak in Merida. I am grateful to all of my research collaborators, on whose work I have freely drawn. Partial financial support for the preparation of this manuscript has been provided at Brown by the US Department of Energy under Grant DE-FG0291ER40688,

\end{document}